\documentclass[12pt]{article}
\usepackage[paper=a4paper,margin=1in]{geometry}
\usepackage{t1enc}      

\usepackage{amsmath}
\usepackage{amsfonts}
\usepackage{amssymb}
\usepackage{amsthm}
\usepackage{graphicx}
\usepackage{mathrsfs}  

\newcommand{\bi}{\bf i}
\newcommand{\bj}{\bf j}

\newcommand{\coker}{\rm coker\,}

\newtheorem{theorem}{Theorem}[section]
\newtheorem{proposition}{Proposition}[section]
\newtheorem{lemma}{Lemma}[section]

\numberwithin{equation}{section}

\begin{document}
\bibliographystyle{unsrt}

\title{Mass, gauge conditions and spectral properties of the 
Sen--Witten and 3-surface twistor operators in closed universes}

\author{ L\'aszl\'o B Szabados \\
Research Institute for Particle and Nuclear Physics \\
H-1525 Budapest 114, P. O. Box 49, Hungary}
\maketitle

\begin{abstract}
A non-negative expression, built from the norm of the 3-surface 
twistor operator and the energy-momentum tensor of the matter 
fields on a spacelike hypersurface, is found which, in the 
asymptotically flat/hyperboloidal case, provides a lower bound 
for the ADM/Bondi--Sachs mass, while on closed hypersurfaces 
coincides with the first eigenvalue of the Sen--Witten operator. 
Also in the closed case, its vanishing is equivalent to the 
existence of non-trivial solutions of Witten's gauge condition. 
Moreover, it is vanishing if and only if the closed data set is 
in a flat spacetime with spatial topology $S^1\times S^1\times 
S^1$. Thus, it provides a positive definite measure of the strength 
of the gravitational field (with physical dimension {\em mass}) 
on closed hypersurfaces, i.e. some sort of the {\em total mass of 
closed universes}. 
\end{abstract}


\section{Introduction}
\label{sec-1}

\subsection{Three problems}
\label{sub-1.1}

\subsubsection{Lower bound for the ADM/Bondi--Sachs masses and 
the mass of closed universes}
\label{sub-1.1.1}

The classical energy positivity proofs guarantee that the total energy 
of asymptotically flat matter+gravity systems, measured both in spatial 
and null infinities (even in the presence of black holes), is bounded 
from below by zero \cite{Wi,Ne,GHHP,ReTo}. Similar but {\em strictly 
positive} lower bound would be provided by the Penrose inequality: the 
total mass could not be less than the irreducible mass associated with 
the black holes \cite{Pe}. (For a review of the status of the Penrose 
inequality, see e.g. \cite{Mars}.) 
Recently, B\"ackdahl and Valiente-Kroon showed by explicit calculation 
\cite{BaKr} that in vacuum, asymptotically flat spacetimes the ADM 
mass can be reexpressed by the norm of the 3-surface twistor operator 
\cite{Tod}, which norm is some form of a geometric invariant of the 
actual spacelike hypersurface. 

This raises the question whether or not the Bondi--Sachs mass can also 
be expressed in an analogous way, perhaps even in the non-vacuum case. 
Or, more generally, whether or not other strictly positive lower 
bounds for the total mass can also be found, even in the absence of 
black holes. It is known that the Hamiltonian structure of general 
relativity dictates that mass/energy-momentum (and other `conserved' 
quantities) should be associated with closed spacelike {\em 2-surfaces} 
(for a review of the various strategies see e.g. \cite{Sz09}). Hence, 
strictly speaking, no such quantity can be expected to be associated 
with a {\em closed spacelike hypersurface} (e.g. in a closed universe). 
Thus, we have the question if a quantity analogous to that behind the 
lower bounds for the ADM and Bondi--Sachs masses could provide certain 
notion of the `total mass' in closed universes.

\subsubsection{Witten-type gauge conditions in closed universes}
\label{sub-1.1.2}

In several specific problems (e.g. in the energy positivity proofs, 
in the Hamiltonian formulation of the theory or in the study of the 
field equations, in particular, in the evolution problems) it is 
desirable to reduce the huge gauge freedom of general relativity. 
Such classical gauge conditions are the Witten \cite{Wi} or Parker 
\cite{Parker} gauges, or Nester's frame gauge condition 
\cite{Ne1,Ne2,Ne3} for spinor or orthonormal frame fields, 
respectively, on spacelike hypersurfaces. Apparently, while in 
the asymptotically flat cases these gauge conditions can be imposed 
and admit non-trivial solutions, explicit calculations indicate that 
the first two cannot be imposed on special, highly symmetric {\em 
closed} spacelike hypersurfaces. 

Thus the question arises naturally whether the Witten type gauge 
conditions can always be imposed at least on generic closed spacelike 
hypersurfaces, and if not, then how those can be generalized. However, 
as far as we know, no such systematic investigation has been devoted 
to this question. On these hypersurfaces Nester's gauge condition (in 
its spinor form) is particularly interesting, because it takes the 
form of a {\em general eigenvalue problem} for a Dirac type operator, 
while the former two require the spinor field to be the eigenspinor 
of a (slightly different) Dirac operator or modified Dirac operator 
with {\em zero eigenvalue}.

\subsubsection{Lower bound for the eigenvalues of the Sen--Witten 
operator in closed universes}
\label{sub-1.1.3}

The eigenvalue problem for Dirac type operators appears in another 
context in geometry and in general relativity. Namely, a promising 
approach to constructing observables of the gravitational field in 
general relativity could be based on the spectral analysis of Dirac 
operators on various submanifolds of the spacetime. For example, the 
eigenvalues of these operators are such gauge invariant objects, which 
are expected to reflect the geometrical properties of the submanifold 
in question, e.g. in the form of some lower bound for the eigenvalues 
in terms of other well known geometrical objects. (For a review of a 
number of related problems in differential geometry, see e.g. 
\cite{Yau}, section IV, pp 685-688.) 

The first who gave such a lower bound in differential geometry was 
Lichnerowicz \cite{L}: he showed, in particular, that on any closed 
Riemannian spin manifold $\Sigma$ with positive scalar curvature 
$\frac{1}{4}\inf\{ R(p)\vert p\in\Sigma\}$ is a lower bound for the 
square of the eigenvalues. However, this bound is not sharp: on a 
metric 2-sphere with radius $r$ the (positive) eigenvalues are 
$\frac{n}{r}$, $n\in{\mathbb N}$, while on metric spheres the bounds 
were expected to be saturated. In fact, in the last two decades such 
{\em sharp} lower bounds were found in terms of the scalar curvature 
\cite{TFr,Hi86,Hi95,TF00}, the more general curvature operator (even 
in the presence of non-trivial boundary conditions) \cite{FaSc} or 
the volume \cite{Ba92,TF00}. In particular, in dimension $m$ the 
sharp lower bound, given by Friedrich \cite{TFr,TF00}, is $\frac{m}
{4(m-1)}\inf\{ R(p)\vert p\in\Sigma\}$. 

To have significance of these results in general relativity we 
should be able to link the bounds to well known concepts of physics, 
e.g. the objects defined in a natural way on a spacelike hypersurface 
$\Sigma$ of a Lorentzian 4-manifold. Such an extension of the pure 
Riemannian geometrical results to spacelike hypersurfaces in 
Lorentzian spin manifolds has in fact been given by Hijazi and Zhang 
\cite{HiZh} using the Sen--Witten operator (acting on Dirac spinors) 
and the technique of Friedrich: the eigenvalues of the Sen--Witten 
operator are bounded from below by a certain average of the 
energy-momentum of the matter fields seen by the observers at rest 
with respect to the hypersurface. 

However, there are non-flat solutions of Einstein's equations even 
in the absence of matter fields, in which case the lower bound of 
\cite{HiZh} is zero, and hence the bound is trivial. Thus, we have 
the question whether or not an even greater, sharp lower bound for 
the eigenvalues can be found which is not zero even in the vacuum 
case. A more ambitious claim is to find an explicit expression for 
the first eigenvalue itself.

\subsection{The aims and results of the paper}
\label{sub-1.2}

The aim of the present paper is to answer the questions above. 
Apparently, these problems seem to be independent, and it is only the 
formalism, e.g. the use of (actually Weyl) spinorial techniques, 
which make them related to each other. However, this is not the case: 
we find a non-negative expression ${\tt M}$, built from the norm of 
the 3-surface twistor operator and the energy-momentum tensor, such 
that (1) in the asymptotically flat and asymptotically hyperboloidal 
cases it provides a lower bound for the ADM {\em and} Bondi--Sachs 
masses, respectively; (2) on {\em closed} spacelike hypersurfaces 
Witten's gauge condition can be imposed if and only if it is 
vanishing, which is also equivalent to the flatness of the spacetime 
with $S^1\times S^1\times S^1$ spatial topology; and (3) it gives the 
first eigenvalue of the Sen--Witten operator on closed spacelike 
hypersurfaces. Thus ${\tt M}$ provides a common generalization of the 
results of \cite{BaKr} and \cite{HiZh}. 
Moreover, since this ${\tt M}$ is some measure of the strength of the 
gravitational `field' (and its physical dimension is {\em mass}, in 
contrast e.g. to the so-called Bel--Robinson energy which does {\em 
not} have the correct dimension \cite{Sz09}), this can also be 
interpreted as some {\em total mass} associated with the whole {\em 
closed} spacelike hypersurface. 

Technically, one of the two key ingredients in our investigations is 
the Sen--Witten identity, given for Weyl spinors in \cite{ReTo}. The 
other is the observation that the decomposition of the derivative of 
a spinor field (with respect to the Sen connection \cite{Se} on the 
spacelike hypersurface $\Sigma$) into its Sen--Witten derivative 
(which is a Dirac operator built from the Sen connection) and the 
3-surface twistor derivative is not only algebraically irreducible, 
but also is an {\em $L_2$--orthogonal decomposition} with respect to 
the natural global $L_2$-scalar product on the space of the spinor 
fields. This decomposition makes it possible to derive a general, 
manifestly non-negative expression, which on the asymptotically 
flat/hyperboloidal hypersurfaces coincides with an appropriate null 
component of the ADM/Bondi--Sachs energy-momentum in the Witten 
gauge, and on closed hypersurfaces it coincides with the norm of 
the Sen--Witten operator. Thus the eigenvalues of this operator are 
given by the general expression of the total energy of the 
matter+gravity systems appearing in Witten's positive energy proof. 
The quantity ${\tt M}$ above is defined to be the infimum of this 
general expression on the space of spinor fields satisfying 
appropriate boundary and normalization conditions. In vacuum the same 
${\tt M}$ trivially gives a lower bound for the eigenvalues of the 
3-surface twistor operator. 

As examples, we calculate the first eigenvalue of the Sen--Witten 
operator on a $t={\rm const}$ hypersurface of the $k=1$ 
Friedmann--Robertson--Walker as well as in the spatially closed Bianchi 
I. cosmological spacetimes. The corresponding eigenspinor is nowhere 
vanishing. This suggests a possible generalization of Witten's gauge 
condition, which is a modification of Nester's condition: the spinor 
field should be the eigenspinor of the Sen--Witten operator with the 
{\em smallest non-negative eigenvalue}. 

In a mathematically complete analysis of the spectral properties of 
the Sen--Witten and the 3-surface twistor operators we should clarify 
some of their functional analytic properties. Since the Sen--Witten 
operator is elliptic, the general theorems and results in the theory 
of elliptic p.d.e. could be applied to it. However, the 3-surface 
twistor operator is only {\em overdetermined elliptic}, and hence 
these theorems cannot be applied to it directly. Also, the Sen--Witten 
operator acting on Dirac spinors is self-adjoint (see e.g. 
\cite{FaSc}), but it is {\em not} when acts on Weyl spinors. Thus, care 
is needed in using results in elliptic p.d.e. theory that are also based 
on the self-adjointness of the elliptic operator. Hence, to have a 
solid functional analytic ground of our investigations, we must carry 
out such a systematic analysis. The key observation is that {\em a 
fundamental estimate for the 3-surface twistor operator can be proven}, 
even though it is not elliptic. This is a consequence of the 
Sen--Witten identity, Einstein's equations and the dominant energy 
condition, i.e. {\em a consequence of our physical assumptions}. 

In section 2 we review the necessary geometrical background, in 
particular the Sen connection and the Sen--Witten identity in the 
form that we use. In section 3, first we derive our fundamental 
identity for the norm of the derivatives of spinor fields. Then we 
show that {\em both} the ADM and Bondi--Sachs energies can be 
expressed by the norm of the 3-surface twistor operator and the 
energy-momentum tensor of the matter fields, and by this expression 
we introduce a non-negative lower bound ${\tt M}$ for them. In 
closed universes the same expression for ${\tt M}$ (but with 
different normalization conditions) is suggested as the total mass. 
It is shown that ${\tt M}=0$ is the necessary and sufficient condition 
of the existence of a non-trivial solution of Witten's gauge condition, 
and that this happens precisely when the spacetime is flat with 
toroidal spatial topology. 

Section 4 is devoted to the eigenvalue problem of the Sen--Witten 
operator. First we discuss the potential difficulties in defining the 
eigenvalue problem for the Sen--Witten operator acting on Weyl spinors. 
Then we show that the first eigenvalue is given by ${\tt M}$, and we 
discuss how this expression is related to the previously given lower 
bounds for the eigenvalues. We conclude this section with a remark on the 
eigenvalue problem for the 3-surface twistor operator and the examples. 

The analysis of the mathematical properties of the Sen--Witten and 
3-surface twistor operators, acting on Weyl spinors, is given in the 
appendix. Here we worked in classical Sobolev spaces over closed 
data sets, but most of the results seem to extend to appropriate 
weighted Sobolev spaces (over asymptotically flat/hyperboloidal data 
sets), as well as to the manifold-with-boundary case when the spinor 
fields are subject to non-trivial (chiral or APS) boundary conditions. 
Since the notations and the formalism of this analysis are the usual 
ones in general relativity (rather than in the p.d.e. theory), we hope 
that this appendix makes the functional analytic techniques available 
for a wider readership in the general relativity community. 

We use the abstract index formalism, and only the boldface indices 
take numerical values. We adopt the sign conventions of \cite{PRI}. 
In particular, the signature of the spacetime metric is $(+,-,-,-)$, 
the curvature and Ricci tensors and the curvature scalar are defined 
by ${}^4R^a{}_{bcd}X^b:=-(\nabla_c\nabla_d-\nabla_d\nabla_c)X^a$, ${}^4
R_{bd}:={}^4R^a{}_{bad}$ and ${}^4R:={}^4R_{ab}g^{ab}$, respectively. 
Then Einstein's equations take the form ${}^4G_{ab}=-\kappa T_{ab}$, 
where $\kappa:=8\pi G$ with Newton's gravitational constant $G$. 
Every manifold and all the geometric structures will be assumed to 
be smooth.


\section{Geometrical preliminaries}
\label{sec-2}

\subsection{Metrics on bundles over $\Sigma$}
\label{sub-2.1}

Let $\Sigma$ be a smooth orientable spacelike hypersurface, $t^a$ 
its future pointing unit normal, and define $P^a_b:=\delta^a_b-t^a
t_b$. This is the orthogonal projection to $\Sigma$, by means of 
which the induced (negative definite) 3-metric is defined by $h
_{ab}:=P^c_aP^d_bg_{cd}$. We assume that the spacetime is space 
and time orientable, at least on an open neighbourhood of $\Sigma$, 
in which case $t^a$ can be (and, in what follows, will be) chosen 
to be globally defined. 

Let $\mathbb{V}^a(\Sigma)$ denote the pull back to $\Sigma$ of the 
spacetime tangent bundle, which decomposes in a unique way to the 
$g_{ab}$-orthogonal direct sum of the tangent bundle $T\Sigma$ and 
the normal bundle of $\Sigma$ spanned by $t^a$. $g_{ab}$ is a 
Lorentzian fiber metric on $\mathbb{V}^a(\Sigma)$, and we call the 
triple $(\mathbb{V}^a(\Sigma),g_{ab},P^a_b)$ the Lorentzian vector 
bundle over $\Sigma$. It is the projection $P^a_b$, as a base point 
preserving bundle map of $\mathbb{V}^a(\Sigma)$ to itself, which 
tells us how the tangent bundle $T\Sigma$ is embedded in $\mathbb{V}
^a(\Sigma)$. Since both $T\Sigma$ and the normal bundle of $\Sigma$ 
in $M$ are globally trivializable, $\mathbb{V}^a(\Sigma)$ is also 
globally trivializable. This implies the existence of a spinor 
structure also. In general, there might be inequivalent spinor 
structures on $\mathbb{V}^a(\Sigma)$, labelled by the elements of the 
first cohomology group of $\Sigma$ with $\mathbb{Z}_2$ coefficients. 
Let us fix such a spinor structure, and let $\mathbb{S}^A(\Sigma)$ 
denote the bundle of 2-component (i.e. Weyl) spinors over $\Sigma$. We 
denote the complex conjugate bundle by $\bar{\mathbb{S}}^{A'}(\Sigma)$. 
As is usual in general relativity (see e.g. \cite{PRI}), we identify 
the Hermitian subbundle of $\mathbb{S}^A(\Sigma)\otimes\bar{\mathbb{S}}
^{A'}(\Sigma)$ with $\mathbb{V}^a(\Sigma)$. Thus, we can convert tensor 
indices to pairs of spinor indices and back freely. 

On the spinor bundle two metrics are defined: the first is the 
natural symplectic metric $\varepsilon_{AB}$, while the other is the 
positive definite Hermitian metric $G_{AB'}:=\sqrt{2}t_{AB'}$. 
(The reason of the factor $\sqrt2$ is that for this definition $G
^{AB'}$, the inverse of $G_{AB'}$ defined by $G^{AB'}G_{BB'}=\delta
^A_B$, is just the contravariant form $\varepsilon^{AC}\varepsilon
^{B'D'}G_{CD'}$ of the Hermitian metric, i.e. the Hermitian and the 
symplectic metrics are compatible.) 
The Hermitian metric defines the ${\mathbb C}$-linear bundle 
isomorphisms $\bar{\mathbb{S}}^{A'}(\Sigma)\rightarrow\mathbb{S}^A
(\Sigma):\bar\lambda^{A'}\mapsto-G^A{}_{A'}\bar\lambda^{A'}$ and 
$\bar{\mathbb{S}}_{A'}(\Sigma)\rightarrow\mathbb{S}_A(\Sigma):\bar
\lambda_{A'}\mapsto G_A{}^{A'}\bar\lambda_{A'}$; as well as 

\begin{equation}
\langle\lambda_A,\phi_A\rangle:=\int_\Sigma G^{AA'}\lambda_A\bar\phi
_{A'}{\rm d}\Sigma, \label{eq:2.1}
\end{equation}
which is a global $L_2$-scalar product on the space $L_2(\Sigma,
\mathbb{S}^A)$ of the (square integrable) spinor fields on 
$\Sigma$. This defines an $L_2$-norm in the standard way: $\Vert
\lambda_A\Vert^2_{L_2}:=\langle\lambda_A,\lambda_A\rangle$. The 
scalar product (\ref{eq:2.1}) and the corresponding norm extend 
in an obvious way to spinor/tensor fields on $\Sigma$ with an 
arbitrary index structure. The conventions above ensure that the 
$L_2$-norm of a {\em spatial} tensor field, say $\Vert T_{ab...}
\Vert_{L_2}$, coincides with that of its spinor form $\Vert T_{AA'
BB'...}\Vert_{L_2}$. 


\subsection{The Sen connection}
\label{sub-2.2}

The intrinsic Levi-Civita covariant derivative operator, defined on 
$T\Sigma$, will be denoted by $D_e$. This will be extended to the 
whole ${\mathbb V}^a(\Sigma)$ by requiring $D_et_a=0$. We introduce 
another connection on ${\mathbb V}^a(\Sigma)$, the so-called Sen 
connection \cite{Se}, by ${\cal D}_a:=P^b_a\nabla_b$. Clearly, both 
$D_e$ and ${\cal D}_e$ annihilate the fiber metric $g_{ab}$, but the 
projection is annihilated only by $D_e$. ($D_e$ is a reduction of 
${\cal D}_e$, and the reduction is made by requiring that the 
projection be annihilated by the covariant derivative operator $D_e$.) 
The extrinsic curvature of $\Sigma$ in $M$ is $\chi_{ab}:={\cal D}_a
t_b=\chi_{(ab)}$. In terms of $D_e$ and the extrinsic curvature the 
action of the Sen derivative on an arbitrary cross section $X^a$ of 
${\mathbb V}^a(\Sigma)$ is given by 

\begin{equation}
{\cal D}_eX^a=D_eX^a+\bigl(\chi_e{}^at_b-t^a\chi_{eb}\bigr)X^b. 
\label{eq:2.2}
\end{equation}
The curvature of ${\cal D}_a$ is defined by the convention $-F^a{}
_{bcd}X^bv^cw^d:=v^c{\cal D}_c(w^d{\cal D}_dX^a)-w^c{\cal D}_c(
v^d{\cal D}_dX^a)-[v,w]^e{\cal D}_eX^a$ for any cross section $X^a$ 
of ${\mathbb V}^a(\Sigma)$ and any $v^c$ and $w^c$ tangent to 
$\Sigma$. This is just the pull back to $\Sigma$ of the spacetime 
curvature 2-form, $F^a{}_{bcd}={}^4R^a{}_{bef}P^e_cP^f_d$, and it can 
be re-expressed as 

\begin{eqnarray}
F_{abcd}\!\!\!\!&=\!\!\!\!&R_{abcd}+\chi_{ac}\chi_{bd}-\chi_{ad}\chi
 _{bc}+ \nonumber \\
\!\!\!\!&+\!\!\!\!&t_a\bigl(D_c\chi_{db}-D_d\chi_{cb}\bigr)-
 t_b\bigl(D_c\chi_{da}-D_d\chi_{ca}\bigr), \label{eq:2.3}
\end{eqnarray}
where $R_{abcd}$ is the curvature tensor of the intrinsic geometry 
of $(\Sigma,h_{ab})$. 

${\cal D}_e$ extends in a natural way to the spinor bundle, and its 
action on a spinor field is 

\begin{equation}
{\cal D}_e\lambda_A=D_e\lambda_A-\chi_{eAA'}t^{A'}{}_B\lambda^B. 
 \label{eq:2.4}
\end{equation}
The commutator of two Sen operators acting on the spinor field 
$\lambda^A$ is

\begin{equation}
\bigl({\cal D}_c{\cal D}_d-{\cal D}_d{\cal D}_c\bigr)\lambda^A=
-F^A{}_{Bcd}\lambda^B-2\chi^e{}_{[c}t_{d]}{\cal D}_e\lambda^A, 
\label{eq:2.5}
\end{equation}
where the curvature $F^A{}_{Bcd}$ is just the pull back to $\Sigma$ 
of the anti-self-dual part ${}^4R^A{}_{Bcd}$ of the spacetime curvature 
2-form , which can also be expressed by the (spinor form of the) 
intrinsic curvature and the extrinsic curvature. 

The Sen--Witten operator, i.e. the Dirac operator built from the Sen 
connection, is defined to be ${\cal D}:C^\infty(\Sigma,{\mathbb S}^A)
\rightarrow C^\infty(\Sigma,\bar{\mathbb S}_{A'}):\lambda^A\mapsto
{\cal D}_{A'A}\lambda^A$, where e.g. $C^\infty(\Sigma,{\mathbb S}^A)$ 
denotes the space of the smooth unprimed, contravariant spinor fields 
on $\Sigma$. Since 

\begin{equation*}
\langle{\cal D}_{A'A}\lambda^A,\bar\phi_{B'}\rangle=\int_\Sigma D
_{AA'}\bigl(\lambda^AG^{A'B}\phi_B\bigr){\rm d}\Sigma+\int_\Sigma
\lambda^AG_{AA'}\bigl({\cal D}^{A'B}\phi_B\bigr){\rm d}\Sigma,
\end{equation*}
the formal adjoint of ${\cal D}$ is ${\cal D}^*:C^\infty(\Sigma,\bar
{\mathbb S}_{A'})\rightarrow C^\infty(\Sigma,{\mathbb S}^A):\bar\phi
_{A'}\mapsto{\cal D}^{AA'}\bar\phi_{A'}$. Thus, writing the latter as 
${\cal D}^*:\bar\phi^{A'}\mapsto-{\cal D}^A{}_{A'}\bar\phi^{A'}$, we 
see that ${\cal D}^*$ is $-1$ times of the complex conjugate of 
${\cal D}$. Therefore, for {\em closed} $\Sigma$ (or on the space 
of the spinor fields for which the first integral on the right is 
vanishing), both ${\cal D}^*{\cal D}:$ $\lambda^A\mapsto{\cal D}
^{AA'}{\cal D}_{A'B}\lambda^B$ and ${\cal D}{\cal D}^*:$ $\bar\phi
_{A'}\mapsto{\cal D}_{A'A}{\cal D}^{AB'}\bar\phi_{B'}$ are formally 
self-adjoint and they are essentially complex conjugate of each 
other. Moreover, since 

\begin{eqnarray}
\langle{\cal D}^{AA'}{\cal D}_{A'B}\lambda^B,\phi^C\rangle\!\!\!\!&
 =\!\!\!\!&\int_\Sigma G_{AA'}\bigl({\cal D}^A{}_{B'}\bar\phi^{B'}
\bigr)\bigl({\cal D}^{A'}{}_B\lambda^B\bigr){\rm d}\Sigma+\nonumber \\
\!\!\!\!&+\!\!\!\!&\int_\Sigma D_{AA'}\Bigl(\bigl({\cal D}^{A'}{}_B
 \lambda^B\bigr)G^A{}_{B'}\bar\phi^{B'}\Bigr){\rm d}\Sigma, 
\label{eq:2.6}
\end{eqnarray}
e.g. for closed $\Sigma$ the operator ${\cal D}^*{\cal D}$ is 
positive: $\langle{\cal D}^{AA'}{\cal D}_{A'B}\lambda^B,\lambda^C
\rangle$ $\geq0$ for every spinor field $\lambda^A$. Note, however, 
that while ${\cal D}^*{\cal D}:C^\infty(\Sigma,\mathbb{S}^A)
\rightarrow C^\infty(\Sigma,\mathbb{S}^A)$ can be extended to be a 
self-adjoint operator on an appropriate subspace of $H_1(\Sigma,
\mathbb{S}^A)$, where $H_k(\Sigma,\mathbb{S}^A)$, $k\in\mathbb{N}$, 
is the $k$th Sobolev space, in subsection \ref{sub-4.1.1} we will 
see that ${\cal D}$ (or, more precisely, ${\rm i}{\cal D}$) is 
{\em not} self-adjoint in the strict sense. It yields a self-adjoint 
operator only on the bundle of the Dirac spinors. Thus, ${\cal D}$ 
could be considered to be self-adjoint on the Weyl spinors in some 
generalized sense.


\subsection{The Sen--Witten identity}
\label{sub-2.3}

Using the commutator (\ref{eq:2.5}), the square of the Sen--Witten 
operator can be written as 

\begin{eqnarray}
{\cal D}_A{}^{A'}{\cal D}_{A'B}\lambda^B\!\!\!\!&=\!\!\!\!&{\cal D}
 _{(A}{}^{A'}{\cal D}_{B)A'}\lambda^B+\frac{1}{2}\varepsilon_{AB}
 {\cal D}_R{}^{R'}{\cal D}_{R'}{}^R\lambda^B= \label{eq:2.7}\\
\!\!\!\!&=\!\!\!\!&-\frac{1}{2}\varepsilon^{A'B'}\bigl({\cal D}
 _{AA'}{\cal D}_{BB'}-{\cal D}_{BB'}{\cal D}_{AA'}\bigr)\lambda^B+
 \frac{1}{2}{\cal D}_e{\cal D}^e\lambda_A=\nonumber \\
\!\!\!\!&=\!\!\!\!&\frac{1}{2}{\cal D}_e{\cal D}^e\lambda_A+
 \frac{1}{2}\varepsilon^{A'B'}F^B{}_{CAA'BB'}\lambda^C+\varepsilon
 ^{A'B'}\chi^e{}_{[a}t_{b]}{\cal D}_e\lambda^B. \nonumber
\end{eqnarray}
The last term can also be written as $\chi^e{}_{AA'}t^{A'}{}_B{\cal 
D}_e\lambda^B$. Using (\ref{eq:2.3}) and the fact that in three 
dimensions the curvature tensor can be expressed by the metric $h
_{ab}$ and the corresponding Ricci tensor and curvature scalar, a 
straightforward computation yields that 

\begin{equation}
\varepsilon^{A'B'}F^B{}_{CAA'BB'}=-\frac{1}{4}\varepsilon_{AC}\bigl(
R+\chi^2-\chi_{de}\chi^{de}\bigr)+\bigl(D_e\chi^e{}_{AA'}-D_{AA'}
\chi\bigr)t^{A'}{}_C. \label{eq:2.8}
\end{equation}
However, the terms on the right hand side are precisely the constraint 
parts of the spacetime Einstein tensor: 

\begin{eqnarray}
\frac{1}{2}\bigl(R+\chi^2-\chi_{ab}\chi^{ab}\bigr)\!\!\!\!&=\!\!\!\!&
 -{}^4G_{ab}t^at^b=\kappa T_{ab}t^at^b=:\kappa \mu, \label{eq:2.9a}\\
\bigl(D_a\chi^a{}_b-D_b\chi\bigr)\!\!\!\!&=\!\!\!\!&-{}^4G_{ae}t^a
 P^e_b=\kappa T_{ae}t^aP^e_b=:\kappa J_b; \label{eq:2.9b} 
\end{eqnarray}
where we used Einstein's field equations. The right hand sides of these 
formulae define the energy density and the spatial momentum density of 
the matter fields, respectively, seen by the observer $t^a$. We will 
assume that the matter fields satisfy the dominant energy condition, 
i.e. $\mu^2\geq\vert J_aJ^a\vert$. Substituting (\ref{eq:2.8}), 
(\ref{eq:2.9a}) and (\ref{eq:2.9b}) into (\ref{eq:2.7}) we obtain 

\begin{eqnarray}
2{\cal D}^{AA'}{\cal D}_{A'B}\lambda^B\!\!\!\!&=\!\!\!\!&{\cal D}_e
 {\cal D}^e\lambda^A+2\chi^{eA}{}_{A'}t^{A'}{}_B{\cal D}_e\lambda^B-
 \nonumber \\
\!\!\!\!&-\!\!\!\!&\frac{1}{2}t_e{}^4G^{ef}t_f\lambda^A+\frac{1}{2}
 t_e\,{}^4G^{ef}P^{AA'}_f2t_{A'B}\lambda^B. \label{eq:2.10}
\end{eqnarray}
This equation is analogous to the Lichnerowicz identity \cite{L} 
(or rather an equation called in the mathematical literature a 
Weitzenb\"ock type equation): the square of the Dirac operator is 
expressed in terms of the Laplacian and the curvature, but here 
${\cal D}_e{\cal D}^e$ is {\em not} the intrinsic Laplacian and, in 
addition, the first derivative of the spinor field also appears on 
the right. Moreover, the curvature in (\ref{eq:2.10}) is not simply 
the scalar curvature, but a genuine tensorial piece of the curvature. 
If, on the other hand, the extrinsic curvature is vanishing, then 
${\cal D}_e$ reduces to the Levi-Civita $D_e$, and (\ref{eq:2.10}) 
reduces to $2D^{AA'}D_{A'B}\lambda^B=D_eD^e\lambda^A+\frac{1}{4}R
\lambda^A$, which is the genuine Lichnerowicz identity for the three 
dimensional intrinsic Dirac operator. 

Contracting (\ref{eq:2.10}) with $t_{AB'}\bar\phi^{B'}$ and using the 
definitions, equation (\ref{eq:2.4}) and the fact that $G^A{}_{B'}G
^{A'}{}_B$ acts on vectors tangent to $\Sigma$ as $-P^a_b$, we obtain 

\begin{eqnarray}
&{}&D_{AA'}\bigl(2t^A{}_{B'}\bar\phi^{B'}{\cal D}^{A'}{}_B\lambda^B
 \bigr)+2t^{AA'}\bigl({\cal D}_{A'B}\lambda^B\bigr)\bigl({\cal D}
 _{AB'}\bar\phi^{B'}\bigr)= \label{eq:2.11} \\
&=\!\!\!\!&D_a\bigl(\bar\phi^{B'}t_{B'B}{\cal D}^a\lambda^B\bigr)-
 t_{AA'} \bigl({\cal D}_e\lambda^A\bigr)\bigl({\cal D}^e\bar\phi
 ^{A'}\bigr)-\frac{1}{2}t^a\,{}^4G_{aBB'}\lambda^B\bar\phi^{B'}. 
 \nonumber
\end{eqnarray}
Writing the total divergences in a different way, we obtain the 
Reula--Tod (or the $SL(2,{\mathbb C})$ spinor) form \cite{ReTo} 
of the Sen--Witten identity: 

\begin{eqnarray}
D_a\bigl(t^{A'B}\bar\phi^{B'}{\cal D}_{BB'}\lambda^A-\bar\phi^{A'}
 t^{AB'}{\cal D}_{B'B}\lambda^B\bigr)\!\!\!\!&+\!\!\!\!&2t^{AA'}
 \bigl({\cal D}_{A'B}\lambda^B\bigr)\bigl({\cal D}_{AB'}\bar\phi
 ^{B'}\bigr)= \nonumber \\
=-t_{AA'}h^{ef}\bigl({\cal D}_e\lambda^A\bigr)\bigl({\cal D}_f\bar
 \phi^{A'}\bigr)\!\!\!\!&-\!\!\!\!&\frac{1}{2}t^a\,{}^4G_{aBB'}
 \lambda^B\bar\phi^{B'}. \label{eq:2.12}
\end{eqnarray}
Clearly, its right hand side is positive definite for $\phi^A=
\lambda^A$ and matter fields satisfying the dominant energy condition. 


\section{Energy-momentum and  gauge conditions}
\label{sec-3}

\subsection{Gravitational energy-momentum}
\label{sub-3.1}

\subsubsection{The key ingredient: the 3-surface twistor operator}
\label{sub-3.1.1}

Using the unitary spinor form ${\cal D}_{EF}:=G_F{}^{E'}{\cal D}
_{E'E}={\cal D}_{(EF)}$ of the Sen derivative operator ${\cal D}_e$ 
(see \cite{Re,Fr}), the decomposition of the derivative ${\cal D}
_e\lambda_A$ into its irreducible parts is 

\begin{eqnarray}
G_F{}^{E'}{\cal D}_{E'E}\lambda_A\!\!\!\!&=\!\!\!\!&{\cal D}_{(EF}
 \lambda_{A)}+\frac{1}{3}\varepsilon_{EA}{\cal D}_{FB}\lambda^B+
 \frac{1}{3}\varepsilon_{FA}{\cal D}_{EB}\lambda^B= \nonumber \\
\!\!\!\!&=\!\!\!\!&{\cal D}_{(EF}\lambda_{A)}+\frac{1}{3}\varepsilon
 _{EA}G_F{}^{K'}{\cal D}_{K'K}\lambda^K+\frac{1}{3}\varepsilon_{FA}
 G_E{}^{K'}{\cal D}_{K'K}\lambda^K= \nonumber \\
\!\!\!\!&=\!\!\!\!&{\cal D}_{(EF}\lambda_{A)}+\frac{1}{3}G_F{}^{E'}
 \bigl(\varepsilon_{EA}\delta^{K'}_{E'}-G_E{}^{K'}G_{E'A}\bigr){\cal 
 D}_{K'K}\lambda^K= \nonumber \\
\!\!\!\!&=\!\!\!\!&{\cal D}_{(EF}\lambda_{A)}+\frac{2}{3}G_F{}^{E'}
 P^{KK'}_{EE'}\varepsilon_{KA}{\cal D}_{K'L}\lambda^L; \label{eq:3.1}
\end{eqnarray}
where the first term on the right, ${\cal D}_{(AB}\lambda _{C)}$, 
is just the 3-surface twistor derivative of the spinor field 
\cite{Tod}, while the second is essentially the Sen--Witten 
operator acting on $\lambda^A$. Indeed, ${\cal D}_{(AB}\lambda _{C)}
=0$ is the purely spatial part in the complete irreducible 3+1 
decomposition of the 1-valence spacetime twistor equation $\nabla
_{A'(A}\lambda_{B)}=0$. A straightforward calculation shows that 
(\ref{eq:3.1}) is in fact the {\em pointwise orthogonal 
decomposition} with respect to $G_{AA'}$, and hence {\em it is 
$\langle\, \cdot,\cdot\,\rangle$--orthogonal also}. Thus, the 
$L_2$-scalar product of the derivative of two spinor fields is 

\begin{equation}
\langle{\cal D}_e\lambda_A,{\cal D}_e\phi_A\rangle=\langle{\cal D}
_{(AB}\lambda_{C)},{\cal D}_{(AB}\phi_{C)}\rangle+\frac{2}{3}\langle
{\cal D}_{A'A}\lambda^A,{\cal D}_{A'B}\phi^B\rangle. \label{eq:3.2} 
\end{equation}
Hence, in particular, the integral of the identity (\ref{eq:2.12}) 
for $\phi^A=\lambda^A$ gives 

\begin{equation}
\Vert{\cal D}_{A'A}\lambda^A\Vert^2_{L_2}=\frac{3}{4}\Vert{\cal D}
_{(AB}\lambda_{C)}\Vert^2_{L_2}+\frac{3}{4\sqrt{2}}\kappa \int_\Sigma t
^aT_{aBB'}\lambda^B\bar\lambda^{B'}{\rm d}\Sigma \label{eq:3.3}
\end{equation}
on a {\em closed} $\Sigma$.


\subsubsection{A lower bound for the ADM and Bondi--Sachs 
masses}
\label{sub-3.1.2}

Suppose for a moment that $\Sigma$ is asymptotically flat or 
asymptotically hyperboloidal and $\lambda^A$ is asymptotically 
constant or satisfies the asymptotic twistor equation at the 
infinity (or infinities) of $\Sigma$, respectively. Let ${}_0
\lambda^A$ denote the asymptotic value of $\lambda^A$ at infinity. 
Let us choose $\phi^A=\lambda^A$ in (\ref{eq:2.12}), and let 
${\tt P}({}_0\lambda{}_0\bar\lambda)$ be $\frac{2}{\kappa}$-times 
of the integral of the total divergence on the left hand side of 
(\ref{eq:2.12}), converted to a 2-surface integral on the boundary 
$\partial\Sigma$ at infinity. Then ${\tt P}({}_0\lambda{}_0\bar
\lambda)$ is just the 2-surface integral of the Nester--Witten 
2-form at infinity, built from $\lambda^A$, and hence is just the 
${}_0\lambda^A{}_0\bar\lambda^{A'}$-component of the ADM or 
Bondi--Sachs energy-momentum, respectively. Hence, from 
(\ref{eq:2.12}), we obtain that 

\begin{equation}
{\tt P}\bigl({}_0\lambda{}_0\bar\lambda\bigr)+\frac{4\sqrt{2}}{3
\kappa}\Vert{\cal D}_{A'A}\lambda^A\Vert^2_{L_2}=\frac{\sqrt{2}}
{\kappa}\Vert{\cal D}_{(AB}\lambda_{C)}\Vert^2_{L_2}+\int_\Sigma t^a
T_{aBB'}\lambda^B\bar\lambda^{B'}{\rm d}\Sigma. \label{eq:3.4}
\end{equation}
This identity is the basis of (probably the simplest) proof of 
the positivity of the ADM and Bondi--Sachs energies, as well as 
of infinitely many different quasi-local energy expressions. 
(For the key ideas and the references in the quasi-local case, 
see e.g. \cite{Sz09}.) The basic idea is that if $\lambda^A$ is 
chosen to be a solution to the Witten equation ${\cal D}_{A'A}
\lambda^A=0$ and satisfying appropriate boundary conditions, 
then the second term on the left hand side is vanishing, and 
hence ${\tt P}({}_0\lambda{}_0\bar\lambda)$ is the sum of two 
manifestly positive definite expressions. 

The gauge condition of Parker \cite{Parker}, formulated in terms 
of Dirac spinors, can also be translated into the language of 
Weyl spinors: since by the dominant energy condition $t^aT_{aBB'}$ 
is a non-negative Hermitian spinor, it has a uniquely determined 
non-negative Hermitian square root $S_{AA'}$ satisfying $G^{AA'}
S_{AB'}S_{A'B}=t^aT_{aBB'}$. (In fact, there is a normalized spin 
frame $\{o^A,\iota^A\}$ such that $\sqrt{2}t^a=o^A\bar o^{A'}+
\iota^A\bar\iota^{A'}$, and the vector $\sqrt{2}v^a:=o^A\bar o^{A'}
-\iota^A\bar\iota^{A'}$ is proportional to the spatial momentum 
density $J_c:=t^aT_{ab}P^b_c$ of the matter fields. In this frame 
$t^aT_{aBB'}=a o_B\bar o_{B'}+b\iota_B\bar\iota_{B'}$, where by the 
dominant energy condition $a,b\geq0$. Then $S_{AA'}:=\sqrt{a}o_A
\bar o_{A'}+\sqrt{b}\iota_A\bar\iota_{A'}$ is the square root 
that we need.) Then Parker's gauge condition is simply ${\cal D}
_{A'A}\psi^A+\gamma S_{A'A}\psi^A=0$ for some real constant 
$\gamma$. Thus, if this equation admits a solution (with given 
boundary conditions) and the constant $\gamma^2$ is chosen to 
be $3\kappa/4\sqrt{2}$, then 

\begin{equation*}
{\tt P}\bigl({}_0\psi{}_0\bar\psi\bigr)=\frac{\sqrt{2}}{\kappa}
\Vert{\cal D}_{(AB}\psi_{C)}\Vert^2_{L_2}. 
\end{equation*}
Now the total energy-momentum of the matter+gravity system is 
represented by the norm of the 3-surface twistor derivative of 
the spinor field $\psi^A$ alone. 

The explicit calculations of B\"ackdahl and Valiente-Kroon showed 
\cite{BaKr} that in asymptotically flat {\em vacuum} spacetimes 
the ADM mass can be given as the $L_2$-norm of the 3-surface 
twistor derivative of (appropriately decaying) asymptotically 
constant spinor fields. The discussion above shows that this 
result can be recovered as a simple consequence of the Sen--Witten 
identity (in the Witten gauge), even in the presence of matter. 
More precisely, {\em the component of both the ADM and Bondi--Sachs 
energy-momenta with respect to any null vector can be written 
as the sum of the $L_2$-norm of the 3-surface twistor operator and 
an energy-momentum term}. In addition, recalling that the 
ADM/Bondi--Sachs energy-momentum is future pointing and timelike 
(or zero) under the conditions of the positive energy theorems, 
for the corresponding ADM/Bondi--Sachs mass ${\tt m}$ we have the 
non-negative lower bound ${\tt m}:=\sqrt{{\tt E}^2-\vert{\tt P}
^{\bi}\vert^2}=\sqrt{({\tt E}-\vert{\tt P}^{\bi}\vert)({\tt E}+
\vert{\tt P}^{\bi}\vert)}\geq{\tt E}-\vert{\tt P}^{\bi}\vert$, 
where ${\tt E}$ and ${\tt P}^{\bi}$, ${\bi}=1,2,3$, are the 
ADM/Bondi--Sachs energy and linear (spatial) momentum, 
respectively, in some global Lorentz frame at infinity, and 
$\vert{\tt P}^{\bi}\vert:=\sqrt{\delta_{\bi\bj}{\tt P}^{\bi}{\tt 
P}^{\bj}}$, the length of the latter. However, ${\tt E}-\vert
{\tt P}^{\bi}\vert$ is just the infimum of ${\tt P}({}_0\lambda
{}_0\bar\lambda)$ on the set of the spinors ${}_0\lambda_A$ for 
which ${}_0t^{AA'}{}_0\lambda_A{}_0\bar\lambda_{A'}=1$, where ${}_0
t^a$ is the timelike basis (unit) vector of the Lorentz frame, 
chosen to be orthogonal to $\Sigma$ at infinity. Therefore, the 
infimum of the right hand side of (\ref{eq:3.4}), 

\begin{equation}
{\tt M}:=\inf\Bigl\{\frac{\sqrt{2}}{\kappa}\Vert{\cal D}_{(AB}
\lambda_{C)}\Vert^2_{L_2}+\int_\Sigma t^aT_{aBB'}\lambda^B\bar\lambda
^{B'}{\rm d}\Sigma\,\Bigr\}, \label{eq:3.5}
\end{equation}
provides a {\em non-negative lower bound for the ADM/Bondi--Sachs 
mass} ${\tt m}$. Here the infimum is taken on the set of spinor 
fields satisfying the appropriate boundary and normalization 
conditions at infinity.

\subsubsection{${\tt M}$ as the mass of closed universes}
\label{sub-3.1.3}

Suppose that $\Sigma$ is a {\em closed} spacelike hypersurface, and 
introduce ${\tt M}$ by (\ref{eq:3.5}), but now the infimum is taken 
on the set of all smooth spinor fields $\lambda_A$ e.g. with norm 
$\Vert\lambda_A\Vert_{L_2}=1$. This provides a measure of the strength 
of the gravitational `field', and we interpret this as the {\em total 
mass of the closed universe}. This interpretation is supported by the 
fact that its physical dimension is mass, and that in the asymptotically 
flat/hyperboloidal case the same formula gives the ADM/Bondi--Sachs 
energy in the Witten gauge. Moreover, in subsection \ref{sub-3.2} we 
will show that ${\tt M}$ is strictly positive, i.e. it is vanishing 
precisely for the trivial data set with toroidal spatial topology. 
This property of ${\tt M}$ is analogous to the rigidity part of the 
positive mass theorems in the asymptotically flat/hyperboloidal case, 
where the vanishing of the total mass implies flatness. 
(Recall that the vanishing of certain spinorial quasi-local {\em 
mass} expressions is equivalent only to {\it pp}-wave Cauchy 
development with pure radiative matter fields. The flatness is 
equivalent to the vanishing of the whole energy-momentum four-vector, 
and not only of its Lorentzian length \cite{Sz93,Sz94,Sz96}.) 

Although the expression between the curly brackets in (\ref{eq:3.5}) 
looks like the $H_1$-Sobolev norm (especially if we write the second 
term in (\ref{eq:3.5}) as $\Vert S_{A'A}\lambda^A\Vert^2_{L_2}$), in 
general it is only a {\em semi-norm}. Indeed, as we will see, this 
`norm' of the Witten spinor on the trivial data set above is vanishing. 

Since ${\tt M}$ was introduced as the infimum of a certain semi-norm 
of {\em smooth} spinor fields, it does not follow {\em a priori} 
that a smooth spinor field $\lambda^A$ with 

\begin{equation}
\Vert{\cal D}_{A'A}\lambda^A\Vert^2_{L_2}=\frac{3}{4\sqrt{2}}
\kappa \, {\tt M}\, \Vert\lambda^A\Vert^2_{L_2}; \label{eq:3.6}
\end{equation}
i.e. with ${\tt M}=\frac{\sqrt{2}}{\kappa}\Vert{\cal D}_{(AB}\lambda
_{C)}\Vert^2_{L_2}+\Vert S_{A'A}\lambda^A\Vert^2_{L_2}$, should exist. 
However, we show that {\em such a smooth spinor field does exist}. 

By (\ref{eq:3.3}) and the definition of ${\tt M}$ it follows that 
$\frac{4\sqrt{2}}{3\kappa}\Vert{\cal D}_{A'A}\chi^A\Vert^2_{L_2}\geq
{\tt M}\Vert\chi^A\Vert^2_{L_2}$ for any smooth spinor field $\chi
^A$. Thus by the definition of infimum there exists a sequence 
$\{\hat\lambda^A_i\}$, $i\in\mathbb{N}$, of smooth spinor fields 
for which $\Vert\hat\lambda^A_i\Vert_{L_2}=1$, and $\hat{L}_i:=
\frac{4\sqrt{2}}{3\kappa}\Vert{\cal D}_{A'A}\hat\lambda^A_i\Vert^2
_{L_2}\rightarrow{\tt M}$ as a monotonically decreasing sequence. 
Since the sequence $\{\hat{L}_i\}$ is bounded, there exists a 
positive constant $K$ such that $\Vert{\cal D}_{A'A}\hat\lambda^A
_i\Vert_{L_2}\leq K$ for any $i\in\mathbb{N}$. Thus by the 
fundamental elliptic estimate for the Sen--Witten operator (see 
Lemma \ref{l:A.2.1} in the appendix) we have that 

\begin{equation*}
\Vert{\cal D}_e\hat\lambda^A_i\Vert_{L_2}<\Vert\hat\lambda^A_i\Vert
_{H_1}\leq\sqrt{2}\Vert{\cal D}_{A'A}\hat\lambda^A_i\Vert_{L_2}+\Vert
\hat\lambda^A_i\Vert_{L_2}\leq\sqrt{2}K+1,
\end{equation*}
i.e. the $L_2$-norm of the derivative of the spinor fields $\hat
\lambda^A_i$, $i\in\mathbb{N}$, as a sequence is bounded with the 
bound $1+\sqrt{2}K$. However, by this boundedness we have the 
freedom to deform the spinor fields such that, for any given $k\in
\mathbb{N}$, the $L_2$-norms of their first $k$ derivatives are also 
bounded; i.e. there exists a sequence $\{\lambda^A_i\}$, $i\in
\mathbb{N}$, of smooth spinor fields such that $\Vert\lambda^A\Vert
_{L_2}=1$, ${L}_i:=\frac{4\sqrt{2}}{3\kappa}\Vert{\cal D}_{A'A}\lambda
^A_i\Vert^2_{L_2}\rightarrow{\tt M}$ as a monotonically decreasing 
sequence, and $\Vert{\cal D}_{e_1}{\cal D}_{e_2}\lambda^A_i\Vert_{L_2}
\leq K_2$, ..., $\Vert{\cal D}_{e_1}\cdots {\cal D}_{e_k}\lambda^A_i
\Vert_{L_2}\leq K_k$ for some  positive constants $K_2$, ..., $K_k$ 
and for all $i\in\mathbb{N}$. Again, the convergence $L_i\rightarrow
{\tt M}$ implies that $\Vert{\cal D}_e\lambda^A_i\Vert_{L_2}\leq K_1$ 
for some constant $K_1>0$ and for all $i\in\mathbb{N}$. Therefore, 

\begin{equation*}
\Vert\lambda^A_i\Vert_{H_k}\leq 1+K_1+\cdots +K_k, \hskip 25pt
\forall i\in\mathbb{N};
\end{equation*}
i.e. the sequence $\{\lambda^A_i\}$ is bounded in the Sobolev space 
$H_k(\Sigma,\mathbb{S}^A)$. Hence, there is a subsequence $\{\lambda
^A_{i_j}\}$, $j\in\mathbb{N}$, which converges to some $\lambda^A_w
\in H_k(\Sigma,\mathbb{S}^A)$ in its {\em weak topology}. But since 
$\{\lambda^A_{i_j}\}$ is bounded and by the Rellich lemma (see appendix 
\ref{sub-A.1}) the injection $H_k(\Sigma,\mathbb{S}^A)\rightarrow L_2
(\Sigma,\mathbb{S}^A)$ is compact, there is a subsequence $\{\lambda
^A_{i_{j_n}}\}$, $n\in\mathbb{N}$, which converges to some $\lambda^A_s
\in L_2(\Sigma,\mathbb{S}^A)$ in the {\em strong topology} of $L_2(
\Sigma,\mathbb{S}^A)$. Since the strong and the weak limits of a 
sequence must coincide, we conclude that we can find a subsequence 
of the sequence $\{\lambda^A_i\}$ which converges strongly to some 
$\lambda^A:=\lambda^A_s=\lambda^A_w\in H_k(\Sigma,\mathbb{S}^A)$. 
Then by the Sobolev lemma (see appendix \ref{sub-A.1}) $H_k(\Sigma,
\mathbb{S}^A)\subset C^{k-2}(\Sigma,\mathbb{S}^A)$ holds, and since 
$k$ is arbitrary, the spinor field $\lambda^A$ is smooth. 

Finally, since both the $L_2$-norm $\Vert\,.\,\Vert_{L_2}:L_2(\Sigma,
\mathbb{S}^A)\rightarrow[0,\infty)$ and the Sen--Witten operator 
${\cal D}:H_1(\Sigma,\mathbb{S}^A)\rightarrow L_2(\Sigma,\bar{
\mathbb{S}}_{A'})$ are continuous, moreover $H_k(\Sigma,\mathbb{S}^A)
\subset H_1(\Sigma,\mathbb{S}^A)$, we have that ${\tt M}=\lim_{n
\rightarrow\infty}L_{i_{j_n}}=\frac{4\sqrt{2}}{3\kappa}\lim_{n
\rightarrow\infty}\Vert{\cal D}_{A'A}\lambda^A_{i_{j_n}}\Vert^2_{L_2}=
\frac{4\sqrt{2}}{3\kappa}\Vert{\cal D}_{A'A}\lambda^A\Vert^2_{L_2}$ 
holds. Therefore, by (\ref{eq:3.3}), this yields ${\tt M}=\frac{
\sqrt{2}}{\kappa}\Vert{\cal D}_{(AB}\lambda_{C)}\Vert^2_{L_2}+\Vert S
_{A'A}\lambda^A\Vert^2_{L_2}$. 

Since $\lambda^A$ in (\ref{eq:3.6}) is smooth, we can rewrite that 
as 

\begin{equation}
\langle2{\cal D}^{AA'}{\cal D}_{A'B}\lambda^B-\frac{3}{2\sqrt{2}}
\kappa\,{\tt M}\,\lambda^A\, ,\, \lambda^A\rangle=0. \label{eq:3.7}
\end{equation}
Thus, either $\lambda^A$ is an eigenspinor of $2{\cal D}^*{\cal D}$ 
with the eigenvalue $\frac{3}{2\sqrt{2}}\kappa{\tt M}$, or $\lambda^A$ 
is orthogonal to $\Delta^A{}_B\lambda^B$, where, for the sake of brevity, 
we introduced the operator $\Delta^A{}_B:=2{\cal D}^{AA'}{\cal D}_{A'B}-
\frac{3}{2\sqrt{2}}\kappa\,{\tt M}\delta^A_B$. We use (\ref{eq:3.7}) in 
subsection \ref{sub-4.2.1} to prove that $\frac{3}{2\sqrt{2}}\kappa
{\tt M}$ is, in fact, the first eigenvalue of the (`square' of the) 
Sen--Witten operator.


\subsection{On Witten type gauge conditions in closed universes}
\label{sub-3.2}

\subsubsection{The local geometry of closed data sets admitting 
a Witten spinor}
\label{sub-3.2.1}

By the existence of a smooth spinor field $\lambda^A$ for which 
(\ref{eq:3.6}) holds, it is clear that {\em Witten's gauge condition 
can be imposed on a closed spacelike hypersurface if and only if ${\tt 
M}=0$}. (The existence of spinor fields satisfying Parker's gauge 
condition can be characterized in a similar way by $\inf\{\Vert{\cal 
D}_{(AB}\lambda_{C)}\Vert_{L_2}\,\vert\,\lambda_A\in C^\infty(\Sigma,
\mathbb{S}_A),\,\Vert\lambda_A\Vert_{L_2}=1\,\}=0$. Thus, the Witten 
and Parker gauge conditions can be imposed only in special geometries.) 

Clearly, if $(\Sigma,h_{ab},\chi_{ab})$ is a data set with flat Sen 
connection ${\cal D}_e$, then the Sen--constant spinor fields solve 
the Witten equation, and hence for such data sets ${\tt M}=0$. In 
the rest of this subsection we determine the geometry of those 
closed data sets $(\Sigma,h_{ab},\chi_{ab})$ which admit non-trivial 
solutions of the Witten equation, i.e. for which ${\tt M}=0$. 

Thus, suppose that $\lambda^A$ is a solution of ${\cal D}_{A'A}\lambda
^A=0$. Then an immediate consequence of (\ref{eq:3.2}), (\ref{eq:3.3}) 
and the dominant energy condition is that $\lambda^A$ is constant with 
respect to ${\cal D}_e$ on $\Sigma$, and that $t^aT_{ab}\lambda^B\bar
\lambda^{B'}=0$. Since by the dominant energy condition $T^a{}_bL^b$ 
must be future pointing and non-spacelike or zero, where $L^a:=
\lambda^A\bar\lambda^{A'}$, its orthogonality to the timelike $t^a$ 
yields that $T_{ab}L^b=0$. Therefore, {\em the algebraic type of the 
energy-momentum tensor of the matter fields must be of pure radiation 
with the wave vector $L^a$}. Thus, it must have the form $T_{ab}=fL_a
L_b$ for some non-negative function $f$. Therefore, the Ricci spinor 
and the curvature scalar of the spacetime geometry is $\Phi_{ABA'B'}=
\frac{1}{2}\kappa f\lambda_A\lambda_B\bar\lambda_{A'}\bar\lambda_{B'}$ 
and $\Lambda=0$, respectively. Hence, at the points of $\Sigma$, the 
anti-self-dual part of the spacetime curvature takes the form $-{}^4
R_{ABCC'DD'}=\Psi_{ABCD}\varepsilon_{C'D'}+\frac{1}{2}\kappa f\lambda_A
\lambda_B\bar\lambda_{C'}\bar\lambda_{D'}\varepsilon_{CD}$. 

Since $\lambda^A$ is constant, it does not have any zero on 
$\Sigma$ (see \cite{Se}). Thus $Z^a:=P^a_b\lambda^B\bar\lambda^{B'}$ 
is a globally defined, nowhere vanishing vector field on $\Sigma$. 
If $\vert Z\vert^2:=-h_{ab}Z^aZ^b$, the positive definite pointwise 
norm of $Z^a$, then by ${\cal D}_e\lambda^A=0$ and the definitions 
it follows that 

\begin{equation}
D_aZ_b=-\vert Z\vert\chi_{ab}. \label{eq:3.8}
\end{equation}
Thus, in particular, $Z_a$ is a {\em closed 1-form}, i.e. {\em 
locally} it has the form $Z_a=D_au$ for some (locally defined) real 
function $u$. Hence, through each point $p\in\Sigma$, there is a 
maximal integral submanifold ${\cal S}_u$ orthogonal to $Z^a$, which 
is a 2-surface in $\Sigma$ given locally by $u={\rm const}$. Let us 
complete the spinor field $\lambda^A$ to be a spin frame $\{\lambda
^A,I^A\}$, normalized by $\lambda_AI^A=1$, such that $N^a:=I^A\bar I
^{A'}$ is orthogonal to the 2-surfaces ${\cal S}_u$ and the complex 
null vectors $M^a:=\lambda^A\bar I^{A'}$ and $\bar M^a:=I^A\bar\lambda
^{A'}$ are tangent to ${\cal S}_u$. Clearly, $M^a$ and $\bar M^a$ 
satisfy the normalization conditions $M_aM^a=\bar M_a\bar M^a=0$ and 
$M_a\bar M^a=-1$; and, if we write $Z^a=-\vert Z\vert v^a$, then $L^a
=\vert Z\vert(t^a-v^a)$ and $2\vert Z\vert N^a=t^a+v^a$. Note, 
however, that while $Z^a$ is globally well defined on $\Sigma$, {\em 
a priori} the vectors $M^a$ and $\bar M^a$ are only locally defined. 

Next we show that the algebraic type of the Weyl spinor at the points 
of the hypersurface is null. Since $\lambda^A$ is constant on $\Sigma$ 
with respect to ${\cal D}_e$, the commutator (\ref{eq:2.5}) gives that 
$\lambda_A{}^4R^A{}_{Bef}P^e_cP^f_d=\lambda_AF^A{}_{Bcd}=0$, i.e. 

\begin{equation*}
\lambda_A{}^4R^A{}_{Bcd}M^c\bar M^d=\lambda_A{}^4R^A{}_{Bcd}Z^cM^d=
\lambda_A{}^4R^A{}_{Bcd}Z^c\bar M^d=0.
\end{equation*}
Substituting the above form of ${}^4R_{ABCC'DD'}$ here and expressing 
$Z^a$ in terms of the null vectors $\lambda^A\bar\lambda^{A'}$ and 
$I^A\bar I^{A'}$ as $Z^a=\frac{1}{2}(\lambda^A\bar\lambda^{A'}-2\vert 
Z\vert^2 I^A\bar I^{A'})$, we find that $\Psi_{ABCD}=\Psi\lambda_A
\lambda_B\lambda_C\lambda_D$ for some complex function $\Psi$ on 
$\Sigma$. Hence, the Weyl spinor is indeed null and $\lambda_A$ is 
its fourfold principal spinor. 

To determine the local geometry of the surfaces ${\cal S}_u$, first 
let us calculate their extrinsic curvature in $\Sigma$. Let us recall 
that $v^a$ is the unit normal to the 2-surfaces along which the 
function $u$ is {\em increasing}. Substituting $Z^a=-\vert Z\vert v^a$ 
into (\ref{eq:3.8}), for the extrinsic curvature we find that $\nu_{ab}
:=\Pi^c_a\Pi^d_bD_cv_d=\chi_{cd}\Pi^c_a\Pi^d_b=:\tau_{ab}$. Here $\Pi^a_b
:=P^a_b+v^av_b$, the orthogonal projection to the 2-surfaces, by means 
of which e.g. the induced metric on ${\cal S}_u$ is $q_{ab}:=h_{cd}
\Pi^c_a\Pi^d_b=-(M_a\bar M_b+\bar M_aM_b)$. Thus the two extrinsic 
curvatures of the 2-surfaces in the spacetime, the $\nu_{ab}$ 
corresponding to their spacelike normal $v_a$ and the $\tau_{ab}$ 
corresponding to the timelike normal $t_a$, coincide. 

Next we calculate the scalar curvature ${}^2{\cal R}$ of the 
intrinsic metric $q_{ab}$ of the 2-surfaces ${\cal S}_u$. Since 
$\nu_{ab}=\tau_{ab}$, the Gauss equation for ${}^2{\cal R}$ in the 
spacetime (see e.g. equation (2.7) in \cite{Sz94a}) gives that 

\begin{eqnarray*}
{}^2{\cal R}\!\!\!\!&=\!\!\!\!&{}^4R_{abcd}q^{ac}q^{bd}-\tau^2+\tau
 _{ab}\tau^{ab}+\nu^2-\nu_{ab}\nu^{ab}={}^4R_{abcd}q^{ac}q^{bd}= \\
\!\!\!\!&=\!\!\!\!&-2F_{abcd}M^a\bar M^bM^c\bar M^d=2\bigl(\bar 
 I^{A'}\bar F_{A'B'cd}\bar\lambda^{B'}-\lambda^AF_{ABcd}I^B\bigr)
 M^c\bar M^d=0.
\end{eqnarray*}
Thus, {\em the maximal integral submanifolds ${\cal S}_u$ are 
intrinsically locally flat 2-surfaces}.


\subsubsection{The global topology of $\Sigma$}
\label{sub-3.2.2}

Let us foliate the spacetime in a neighbourhood of $\Sigma$ by 
spacelike hypersurfaces obtained from $\Sigma$ by Lie dragging it 
along its own unit timelike normal $t^a$, and consider the Weyl 
neutrino equation in its 3+1 form with respect to this foliation: 

\begin{equation*}
0=\nabla_{A'A}\lambda^A=t^e\bigl(\nabla_e\lambda^A\bigr)t_{AA'}
+{\cal D}_{A'A}\lambda^A.
\end{equation*}
It is known that this equation admits a well posed initial value 
formulation, and hence, for any given initial spinor field 
$\lambda^A$ on $\Sigma$, it has a unique solution (at least in 
a neighbourhood $\Sigma\times(-\epsilon,\epsilon)$ of $\Sigma$ 
for some $\epsilon>0$). In particular, it has a solution for the 
initial spinor fields satisfying Witten's equation. For such 
spinor fields $t^e\nabla_e\lambda^A=0$ holds, and we show that {\em 
this spinor field is also constant with respect to the spacetime 
connection}. Since ${\cal D}_{A'A}\lambda^A=0$ on $\Sigma$ implies 
${\cal D}_e\lambda^A=0$, we have that on $\Sigma$ 

\begin{eqnarray*}
t^e\nabla_e\bigl({\cal D}_{A'A}\lambda^A\bigr)\!\!\!\!&=\!\!\!\!&
 t^e\nabla_e\bigl(-t_{A'A}t^f\bigr)\nabla_f\lambda^A+t^eP^f_{A'A}
 \nabla_e\nabla_f\lambda^A= \\
\!\!\!\!&=\!\!\!\!&t^eP^f_{A'A}\bigl(\nabla_e\nabla_f-\nabla_f
 \nabla_e\bigr)\lambda^A+t^e{\cal D}_{A'A}\bigl(\nabla_e\lambda^A
 \bigr)= \\
\!\!\!\!&=\!\!\!\!&t^eP^f_{A'A}\bigl(\Psi\lambda^A\lambda_B\lambda
 _E\lambda_F\varepsilon_{E'F'}+\frac{1}{2}\kappa f\lambda^A\lambda
 _B\bar\lambda_{E'}\bar\lambda_{F'}\varepsilon_{EF}\bigr)\lambda^B
 =0.
\end{eqnarray*}
Thus, the spinor field $\lambda^A$ satisfies the Witten equation 
on the neighbouring leaves of the foliation, and hence it is also 
constant with respect to the Sen connection there. However, this, 
together with $t^e\nabla_e\lambda^A=0$, is equivalent to $\nabla_e
\lambda^A=0$. 

Since $\lambda^A$ is constant with respect to the spacetime 
connection, $\nabla_aL_b=0$ also holds, i.e. {\em $L_a$ is a 
constant null vector field}. Thus the spacetime has a {\it 
pp}-wave geometry with the wave vector $L^a$. Hence, $L^a$ is the 
tangent of the null geodesic generators of null hypersurfaces 
${\cal L}$. The intersection of these null hypersurfaces with 
$\Sigma$, ${\cal S}:={\cal L}\cap\Sigma$, gives just the maximal 
integral submanifolds of the previous subsection. Therefore, 
the 2-surfaces ${\cal S}$ are {\em globally well defined closed 
orientable surfaces}. Since the induced metric on these 2-surfaces 
is locally flat, by the Gauss--Bonnet theorem we obtain that {\em 
the topology of the 2-surfaces ${\cal S}$ is torus}: ${\cal S}
\approx S^1\times S^1$. 

The null hypersurfaces ${\cal L}$ can be labelled {\em locally} by 
the value of the function $u$ for which ${\cal L}\cap\Sigma={\cal 
S}_u$ holds. This yields an extension of $u$ from open domains in 
$\Sigma$ to open domains in $\Sigma\times(-\epsilon,\epsilon)$. 
Since $Z_a=D_au$ is nowhere vanishing on $\Sigma$, this $u$ does 
not have any critical point, i.e. locally $u$ provides a 
parametrization of the global foliation of $\Sigma$ by the 
intrinsically flat toroidal 2-surfaces ${\cal S}_u$. 

We show that {\em $\Sigma$ is homeomorphic to the three-torus}: 
$\Sigma\approx S^1\times S^1\times S^1$. As the first step, we 
show that it is a fiber bundle over $S^1$ with typical fiber $S^1
\times S^1$. Let $B:=\Sigma/{\cal S}$, the set of the 2-surfaces 
${\cal S}_u$ in $\Sigma$, and let $\pi:\Sigma\rightarrow B$ be the 
natural projection. Thus the points $[p]\in B$, $p\in\Sigma$, are 
equivalence classes of points of $\Sigma$, where $p$ 
and $q$ are considered to be equivalent if $p,q\in{\cal S}_u$ for 
some $u$. If $(x^1,x^2)$ are local coordinates in a neighbourhood 
of $p\in{\cal S}_u$ on ${\cal S}_u$, then by the non-vanishing of 
$Z^a$ these coordinates can be extended along the integral curves 
of $Z^a$ by $Z^aD_ax^1=Z^aD_ax^2=0$ onto the neighbouring surfaces. 
Thus $(u,x^1,x^2)$ forms a local coordinate system on $\Sigma$, 
while $u$ is a local coordinate on $B$. Then, in these coordinates, 
the projection $\pi$ is simply $(u,x^1,x^2)\mapsto u$, which is 
clearly smooth, and $B$ is a one-dimensional manifold. In addition, 
since $Z^a$ has no zeros, short enough open intervals in $B$ are 
obviously local trivialization domains for $\Sigma$. Hence $\pi:
\Sigma\rightarrow B$ is a smooth fiber bundle with typical fiber 
$S^1\times S^1$ and $B$ is a one-dimensional smooth manifold. 
Finally, since $\Sigma$ is compact and $\pi$ is surjective and 
continuous, $B$ must also be compact, i.e. $B$ is topologically 
$S^1$. 

To show that $\Sigma$, as a bundle, is globally trivial, let us 
cover the base manifold $B\approx S^1$ by two overlapping local 
trivialization domains. These overlap in two disjoint intervals, 
say $U_1$ and $U_2$. The transition function on one, say $\psi_1:
U_1\rightarrow{\rm Diff}(S^1\times S^1)$, can be chosen to be the 
constant map $u\mapsto{\rm Id}$, assigning the identity 
diffeomorphism of $S^1\times S^1$ to every $u\in U_1$. Then the 
different bundles are in a one-to-one correspondence with the 
homotopy classes of the other transition function $\psi_2:U_2
\rightarrow{\rm Diff}(S^1\times S^1)$. Since $U_2$ is a connected 
interval, there are two such homotopy classes: one is the homotopy 
class of the maps into the orientation preserving component of 
${\rm Diff}(S^1\times S^1)$, while the other is the homotopy class 
of the maps into the orientation changing component. Since, however, 
the second yields a bundle whose total space is not orientable 
while $\Sigma$ is orientable, we conclude that $\psi_2:U_2
\rightarrow{\rm Diff}(S^1\times S^1)$ is homotopic with the constant, 
identity map $u\mapsto{\rm Id}$. The corresponding bundle is 
therefore the globally trivial one: $\Sigma\approx S^1\times S^1
\times S^1$. (For the classification of bundles over $n$--spheres, 
and also for the related ideas, see e.g. \cite{St}, section 18, 
pp 96.)  


\subsubsection{The line element}
\label{sub-3.2.3}

To determine the form of the spacetime metric (on a neighbourhood 
of $\Sigma$), we can adapt the strategy of \cite{Sz96} (developed 
originally for the quasi-local case) to the present closed case. 
The main points of this analysis are as follows: 

First, let us combine the surface coordinates $(x^1,x^2)$ into the 
complex coordinate $\zeta=\frac{1}{\sqrt{2}}(x^1-{\rm i}x^2)$ on 
the 2-surfaces, and then complete the local spatial coordinate 
system $(u,\zeta,\bar\zeta)$ on $\Sigma$ to a local coordinate 
system $(u,\zeta,\bar\zeta,v)$ in a neighbourhood of $\Sigma$. 
(Since topologically $\Sigma$ is a 3-torus, the domain of the 
spatial coordinates is finite, say $0\leq x^1,x^2<2\pi$ and $0\leq 
u<u_+$.) Here the coordinate $v$ is the affine parameter along the 
future pointing null geodesic generators of the null hypersurfaces 
${\cal L}$, measured from $\Sigma$. In these coordinates $L^a=g^{ab}
\nabla_bu=(\frac{\partial}{\partial v})^a$ and $M^a=(\frac{\partial}
{\partial\bar\zeta})^a$, and $M^a$ and $\bar M^a$ are already {\em 
globally defined}. 

Since $L^a=(\frac{\partial}{\partial v})^a$ is a Killing vector, 
neither $\Psi$ nor $f$, the only {\em a priori} not zero components 
of the curvature, depends on $v$. 
In the coordinates $(u,\zeta,\bar\zeta,v)$ the line element of the 
spacetime metric takes the form $ds^2=2Hdu^2+2du\,dv+2(Gd\zeta+\bar 
Gd\bar\zeta)du-2d\zeta\,d\bar\zeta$, where $H=H(u,\zeta,\bar\zeta)$ 
is a real and $G=G(u,\zeta,\bar\zeta)$ is a complex function. Both 
are globally defined on $\Sigma$. Since $0=I^A{\cal D}_e\lambda_A=
\lambda^A{\cal D}_eI_A=M^a{\cal D}_e\bar M_a$, and its right hand 
side is just a Christoffel symbol that can be calculated from the 
line element, it yields that $(\partial G/\partial\bar\zeta)=
(\partial\bar G/\partial\zeta)$. Recalling that in the complex 
coordinates the flat Laplacian on ${\cal S}_u$ is $\partial^2/
\partial\zeta\partial\bar\zeta$, this implies that, apart from a 
purely $u$-dependent additive term, there is a uniquely determined 
globally defined {\em real} function $V=V(u,\zeta,\bar\zeta)$ such 
that $(\partial^2V/\partial\zeta\partial\bar\zeta)=-(\partial G/
\partial\bar\zeta)$. Therefore, $G_0:=G+(\partial V/\partial\zeta)$ 
is a globally defined {\em anti-holomorphic function}, and hence by 
Liouville's theorem it does not depend on $\zeta$ and $\bar\zeta$; 
i.e. $G_0=G_0(u)$. (In the language of differential forms $G=G_0-
(\partial V/\partial\zeta)$ is just the Hodge decomposition of the 
closed 1-form on ${\cal S}_u$ with the components $(\frac{1}{\sqrt{2}}
(G+\bar G),-\frac{\rm i}{\sqrt{2}}(G-\bar G))$ into the sum of a 
harmonic form, represented by $G_0$, and an exact form.) 

Finally, by the transformation $(u,\zeta,\bar\zeta,v)\mapsto(u,\zeta,
\bar\zeta,v+V(u,\zeta,\bar\zeta))$ of the coordinates we can change 
the form of the line element such that in the new form $G_0$ appears 
in place of $G$, and the new $H$ is also denoted by $H$. In these new 
coordinates for the only non-zero components of the Weyl and Ricci 
spinors, respectively, we have the expressions 

\begin{equation}
\Psi=\frac{\partial^2H}{\partial\zeta^2}, 
\hskip 40pt
\frac{1}{2}\kappa f=\frac{\partial^2H}{\partial\zeta\partial\bar
\zeta}.  \label{eq:3.9}
\end{equation}
However, the second is a Poisson equation for $H$ on the {\em 
closed} ${\cal S}_u$, which can have a solution only if the 2-surface 
integral of its source term, $\frac{1}{2}\kappa f$, is vanishing. 
Thus by $f\geq0$ (i.e. the dominant energy condition) this implies 
$f=0$, and hence $H$ must be a real harmonic function on ${\cal S}_u$. 
Hence, $H=H(u)$ and, by the first equation of (\ref{eq:3.9}), $\Psi
=0$ follows. Therefore, all the not {\em a priori} vanishing 
components of the spacetime curvature tensor have also been shown to 
be vanishing. Thus the spacetime is locally flat with toroidal spatial 
topology.


\subsection{On Nester's gauge condition in closed universes}
\label{sub-3.3}

By the results of subsection \ref{sub-3.2} Witten's gauge condition 
cannot be imposed in non-flat closed universes. One such alternative 
condition might be that of Nester \cite{Ne1,Ne2,Ne3}. (For a recent 
discussion of the full gauge condition, namely the additional 
requirement of the non-vanishing of the solutions of this equation, 
see e.g. \cite{FNSz}.) In the present Weyl spinor formalism the 
equation underlying Nester's gauge condition is 

\begin{equation}
{\rm i}G_A{}^{B'}D_{B'B}\lambda^B=\frac{1}{\sqrt{2}}\beta\lambda_A
\label{eq:3.10}
\end{equation}
for {\em some} real constant $\beta$. In terms of the Sen connection 
this takes the form ${\cal D}_{A'A}\lambda^A=\frac{1}{2\sqrt{2}}(\chi
-2{\rm i}\beta)G_{A'A}\lambda^A$. Substituting this into (\ref{eq:3.3}) 
and using the definition of ${\tt M}$ we obtain that 

\begin{equation}
\beta^2\geq\frac{3}{2\sqrt{2}}\kappa{\tt M}-\frac{1}{4}\sup\{\chi
^2(p)\vert\,p\in\Sigma\,\}. \label{eq:3.11}
\end{equation}
Thus Nester's gauge condition can be imposed only for those $\beta$ 
which satisfy the inequality (\ref{eq:3.11}). However, this gauge 
condition in the form (\ref{eq:3.10}) is an {\em eigenvalue problem} 
for the Dirac operator built from the {\em intrinsic Levi-Civita 
covariant derivative operator $D_e$}. Thus the results on the 
eigenvalue problem (\ref{eq:3.10}) can be obtained from the 
corresponding results for the Sen--Witten operator, derived in the 
following section, by the formal substitution $\chi_{ab}=0$.


\section{The eigenvalue problem}
\label{sec-4}

\subsection{The eigenvalue problem for the Sen--Witten 
operator}
\label{sub-4.1}

\subsubsection{An attempt with Weyl spinors}
\label{sub-4.1.1}

According to the general theory of spinors (see e.g. the appendix of 
\cite{PRII}) in three dimensions the spinors have two components, 
moreover the Sen--Witten operator maps cross sections of ${\mathbb 
S}^A(\Sigma)$ to cross sections of the complex conjugate bundle $\bar
{\mathbb S}_{A'}(\Sigma)$, it seems natural to define the eigenvalue 
problem by the unitary spinor form \cite{Re,Fr} of ${\cal D}_{AA'}$, 
according to 

\begin{equation}
{\rm i}{\cal D}_A{}^B\psi_B=-\frac{1}{\sqrt{2}}\beta\psi_A. 
\label{eq:4.1}
\end{equation}
(The choice for the apparently ad hoc coefficient $-1/\sqrt{2}$ in 
front of the eigenvalue $\beta$ yields the compatibility with the 
known standard results in special cases. See also (\ref{eq:3.10}).) 
However, it is desirable that the Hermitian metric be compatible 
with the connection in the sense that ${\cal D}_eG_{AA'}=0$. 
Unfortunately, since ${\cal D}_eG_{AA'}$ is $\sqrt2$-times the 
extrinsic curvature of $\Sigma$, in general this requirement cannot 
be satisfied. As a consequence, in general the eigenvalue $\beta$ 
is {\em not} real. In fact, a straightforward calculation (by 
elementary integration by parts) gives that 

\begin{equation}
\beta\Vert\psi_A\Vert^2_{L_2}=\bar\beta\Vert\psi_A\Vert^2_{L_2}+{\rm i}
\int_\Sigma\chi G_{AA'}\psi^A\bar\psi^{A'}{\rm d}\Sigma+{\rm i}\sqrt{2}
\int_\Sigma D_{AA'}\bigl(\psi^A\bar\psi^{A'}\bigr){\rm d}\Sigma. 
\label{eq:4.2}
\end{equation}
This implies that, even if $\Sigma$ is closed (which will be 
assumed in the rest of this paper), the imaginary part of $\beta$ 
is proportional to the integral of the mean curvature $\chi$ 
weighted by the pointwise norm $G_{AA'}\psi^A\bar\psi^{A'}$, which 
is not zero in general. This indicates that the operator ${\rm i}
{\cal D}_A{}^B:C^\infty(\Sigma,\mathbb{S}_A)\rightarrow C^\infty(
\Sigma,\mathbb{S}_A)$ is {\em not} even formally self-adjoint, and 
hence it cannot be made self-adjoint in the strict sense. In fact, 
for any two smooth spinor fields $\lambda_A$ and $\mu_A$ we have 
$\langle{\rm i}{\cal D}_A{}^B\lambda_B,\mu_A\rangle=\langle\lambda
_A,{\rm i}{\cal D}_A{}^B\mu_B\rangle-\frac{\rm i}{\sqrt{2}}\int
_\Sigma\chi G_{AA'}\lambda^A\bar\mu^{A'}{\rm d}\Sigma$. 


\subsubsection{The definition with Dirac spinors}
\label{sub-4.1.2}

This difficulty raises the question whether we can find a slightly 
different definition of the eigenvalue problem for the Sen--Witten 
operator yielding {\em real} eigenvalues. To motivate this, observe 
that although the base manifold $\Sigma$ is only three dimensional, 
the connection ${\cal D}_e$ is four dimensional in its spirit, as 
originally it is defined on the Lorentzian vector bundle ${\mathbb 
V}^a(\Sigma)$. Since its fibers are four dimensional, the 
corresponding spinors are the four component Dirac spinors. Hence 
we should define the eigenvalue problem for the Sen--Witten operator 
in terms of the Dirac spinors. 

Recall that a Dirac spinor $\Psi^\alpha$ is a pair of Weyl spinors 
$\lambda^A$ and $\bar\mu^{A'}$, written them as a column vector 

\begin{equation}
\Psi^\alpha=\left(\begin{array}{cc}\lambda^A \\  
                          \bar\mu^{A'}\end{array}\right) 
\label{eq:4.3}
\end{equation}
and adopting the convention $\alpha=A\oplus{A'}$, $\beta=B\oplus{B'}$ 
etc. Its derivative ${\cal D}_e\Psi^\alpha$ is the column vector 
consisting of ${\cal D}_e\lambda^A$ and ${\cal D}_e\bar\mu^{A'}$. If 
Dirac's $\gamma$-`matrices' are denoted by $\gamma^\alpha_{e\beta}$, 
then one can consider the eigenvalue problem 

\begin{equation}
{\rm i}\gamma^\alpha_{e\beta}{\cal D}^e\Psi^\beta=\alpha\Psi^\alpha.  
\label{eq:4.4}
\end{equation}
Explicitly, with the representation 

\begin{equation}
\gamma^\alpha_{e\beta}=\sqrt{2}\left(\begin{array}{cc}
          0&\varepsilon_{E'B'}\delta^A_E \\
      \varepsilon_{EB}\delta^{A'}_{E'}&0 \end{array}\right) 
\label{eq:4.5}
\end{equation}
(see e.g. \cite{PRI}, pp 221), this is just the pair of equations 

\begin{equation}
{\rm i}{\cal D}_{A'}{}^A\lambda_A=-\frac{\alpha}{\sqrt2}\bar\mu_{A'}, 
\hskip 20pt
{\rm i}{\cal D}_A{}^{A'}\bar\mu_{A'}=-\frac{\alpha}{\sqrt2}\lambda_A.
\label{eq:4.6}
\end{equation}
These imply that both the unprimed and the primed Weyl spinor parts 
of $\Psi^\alpha$ are eigenspinors of ${\cal D}^*{\cal D}$ and its 
complex conjugate, respectively, with the {\em same} eigenvalue: 

\begin{equation}
2{\cal D}^{AA'}{\cal D}_{A'B}\lambda^B=\alpha^2\lambda^A, 
\hskip 20pt
2{\cal D}^{A'A}{\cal D}_{AB'}\bar\mu^{B'}=\alpha^2\bar\mu^{A'}.
\label{eq:4.7}
\end{equation}
Then by (\ref{eq:2.6}) $0\leq2\langle{\cal D}^{AA'}{\cal D}_{A'B}
\lambda^B,\lambda^C\rangle=\alpha^2\Vert\lambda^A\Vert^2_{L_2}$, 
i.e. {\em the eigenvalues $\alpha$ are real}. Conversely, if 
the pair $(\alpha^2,\lambda^A)$ is a solution of the eigenvalue 
problem for $2{\cal D}^*{\cal D}$ with nonzero real $\alpha$, 
then $(\pm\alpha,\Psi^\alpha_\pm)$ with $\bar\mu^{A'}:=\mp(\sqrt{2}
/\alpha){\rm i}{\cal D}^{A'A}\lambda_A$ are solutions of the 
eigenvalue problem (\ref{eq:4.4}). 

By (\ref{eq:4.6}) $\Psi^\alpha=(\lambda^A,\bar\mu^{A'})$ is a Dirac 
eigenspinor with eigenvalue $\alpha$ precisely when $(\lambda^A,-
\bar\mu^{A'})$ is a Dirac eigenspinor with eigenvalue $-\alpha$. 
In the language of Dirac spinors this is formulated in terms of 
the chirality, represented by the so-called `$\gamma_5$-matrix', 
denoted here by 

\begin{equation}
\eta^\alpha{}_\beta:=\frac{1}{4!}\varepsilon^{abcd}\gamma^\alpha
_{a\mu}\gamma^\mu_{b\nu}\gamma^\nu_{c\rho}\gamma^\rho_{d\beta}=
{\rm i}\left(\begin{array}{cc} \delta^A_B&0\\ 
        0&-\delta^{A'}_{B'}\end{array}\right) 
\label{eq:4.8}
\end{equation}
(see appendix II. of \cite{PRII}). Since this is anti-commuting with 
$\gamma^\alpha_{e\beta}$, from (\ref{eq:4.4}) we obtain that ${\rm i}
\gamma^\alpha_{e\mu}{\cal D}^e(\eta^\mu{}_\beta\Psi^\beta)=-\alpha
(\eta^\alpha{}_\beta\Psi^\beta)$. Thus if $\Psi^\alpha$ is a Dirac 
eigenspinor with eigenvalue $\alpha$, then, in fact, $\eta^\alpha{}
_\beta\Psi^\beta$ is a Dirac eigenspinor with eigenvalue $-\alpha$. 
On the other hand, if there are Dirac eigenspinors with definite 
chirality, then they  belong to the kernel of the Sen--Witten 
operator. Indeed, Dirac spinors with definite chirality have the 
structure either $(\lambda^A,0)$ or $(0,\bar\mu^{A'})$, which, by 
(\ref{eq:4.6}), yields that ${\cal D}_{A'A}\lambda^A=0$ or ${\cal D}
_{AA'}\bar\mu^{A'}=0$, respectively. Therefore, this notion of 
chirality cannot be used to decompose the space of the eigenspinors 
with given eigenvalue. Its role is simply to take a Dirac eigenspinor 
with eigenvalue $\alpha$ to a Dirac eigenspinor with eigenvalue 
$-\alpha$. 

By the reality of the eigenvalues both the unprimed spinor part 
$\lambda^A$ and the complex conjugate of the primed spinor part 
$\bar\mu^{A'}$ of $\Psi^\alpha$ are eigenspinors of $2{\cal D}^*{\cal 
D}$ with the same eigenvalue $\alpha^2$. This raises the question 
if the eigenvalue problem can be restricted by $\lambda^A=\mu^A$, 
i.e. by requiring the Dirac eigenspinors $\Psi^\alpha$ to be 
Majorana spinors also. However, (\ref{eq:4.6}) implies that in 
this case $\alpha$ would have to be purely imaginary or zero, i.e. 
the Sen--Witten operator does not have genuine, non-trivial 
Majorana eigenspinors. 

Finally, we consider the special case in which the extrinsic curvature 
is vanishing. In this case ${\cal D}_e=D_e$, and let us consider 
the eigenvalue problem defined by (\ref{eq:4.1}). Then ${\rm i}G_A
{}^{A'}D_{A'}{}^B({\rm i}G_B{}^{B'}D_{B'}{}^C\psi_C)=\frac{1}{2}\beta
^2\psi_A$. However, by $D_eG_{AA'}=0$ we can write 

\begin{eqnarray*}
\beta^2\psi_A\!\!\!\!&=\!\!\!\!&-2G_A{}^{A'}G^B{}_{B'}D_{A'B}\bigl(
 D^{B'C}\psi_C\bigr)=-2G_A{}^{A'}G_{A'}{}^BD_{BB'}\bigl(D^{B'C}\psi
 _C\bigr)= \\
\!\!\!\!&=\!\!\!\!&-2G_{AA'}G^{A'B}\bigl(D_B{}^{B'}D_{B'}{}^C\psi_C
 \bigr)=-2D_A{}^{A'}D_{A'}{}^B\psi_B.
\end{eqnarray*}
Thus the pair $(\beta,\psi^A)$ is a solution of the eigenvalue 
problem for $2D^*D$, and hence we may write $\beta=\alpha$ and 
$\psi^A=\lambda^A$. Then $\alpha\bar\mu_{A'}=-{\rm i}\sqrt{2}D_{A'}
{}^A\lambda_A={\rm i}\sqrt{2}G_{A'}{}^AG_A{}^{B'}D_{B'}{}^B\lambda_B=
{\rm i}\sqrt{2}G_{A'}{}^A(\frac{1}{\sqrt2}{\rm i}\alpha\lambda_A)
=\alpha G_{A'A}\lambda^A$; i.e. the primed spinor part $\bar\mu
_{A'}$ of the Dirac eigenspinor is just $G_{A'A}\lambda^A$. Hence, 
in the special case of the vanishing extrinsic curvature the 
eigenvalue problems (\ref{eq:4.1}) and (\ref{eq:4.4}) coincide. 

Therefore, in the general case, there is no way to reduce the 
eigenvalue problem (\ref{eq:4.4}) to a simpler one with first 
order operator and only with a single Weyl spinor. We must study 
either the eigenvalue problem with the first order Sen--Witten 
operator but with Dirac spinors, or with a single Weyl spinor 
but with the second order operator ${\cal D}^*{\cal D}$. In the 
present paper we choose the second strategy, and, for the sake 
of completeness, in the appendix we prove and summarize the key 
theorems on the functional analytic properties of ${\cal D}^*
{\cal D}$. In particular, ${\cal D}^*{\cal D}$ is a non-negative, 
self-adjoint operator with a pure discrete spectrum. 


\subsection{Lower bounds for the eigenvalues of the Sen--Witten 
and 3-surface twistor operators}
\label{sub-4.2}

\subsubsection{${\tt M}$ as the first eigenvalue of ${\cal D}^*{\cal 
D}$}
\label{sub-4.2.1}

Suppose that $\lambda^A$ is an eigenspinor of $2{\cal D}^*{\cal D}$ 
with eigenvalue $\alpha^2$. Then since we assumed that $\Sigma$ is 
closed, (\ref{eq:2.6}), (\ref{eq:2.12}) and (\ref{eq:4.7}) yield 
that 

\begin{equation}
\alpha^2\Vert\lambda^A\Vert^2_{L_2}=\Vert{\cal D}_e\lambda^A\Vert^2
_{L_2}+\frac{\kappa}{\sqrt{2}}\int_\Sigma t^aT_{aBB'}\lambda^B\bar
\lambda^{B'}\,{\rm d}\Sigma,  \label{eq:4.9}
\end{equation}
implying a lower bound for the eigenvalue $\alpha^2$: 

\begin{equation}
\alpha^2\geq\frac{\kappa}{\sqrt{2}\Vert\lambda^A\Vert^2_{L_2}}
\int_\Sigma t^aT_{aBB'}\lambda^B\bar\lambda^{B'}\,{\rm d}\Sigma
\geq\frac{1}{2}\kappa\inf\frac{\int_\Sigma t^aT_{ab}l^b\,{\rm d}
\Sigma}{\int_\Sigma t_el^e\,{\rm d}\Sigma}. \nonumber
\end{equation}
Here the infimum is taken on the set of the smooth, future pointing 
null vector fields $l^a$ on $\Sigma$. However, this bound is certainly 
{\em not} sharp: in the special case of the vanishing extrinsic 
curvature the nominator is the integral of $\frac{1}{2}Rt_al^a$ (see 
equations (\ref{eq:2.9a})-(\ref{eq:2.9b})), yielding Lichnerowicz's 
bound $\frac{1}{4}\inf\{ R(p)\vert p\in\Sigma\}$ instead of 
Friedrich's sharp bound $\frac{3}{8}\inf\{ R(p)\vert p\in\Sigma\}$. 

To find the sharp bound, let us use (\ref{eq:3.2}) with $\phi^A=
\lambda^A$ in (\ref{eq:4.9}). We obtain 

\begin{equation}
\alpha^2\Vert\lambda^A\Vert^2_{L_2}=\frac{3}{2}\Vert{\cal D}_{(AB}
\lambda_{C)}\Vert^2_{L_2}+\frac{3}{2\sqrt{2}}\kappa\int_\Sigma t^a
T_{aBB'}\lambda^B\bar\lambda^{B'}\,{\rm d}\Sigma, \label{eq:4.10}
\end{equation}
from which the lower bound 

\begin{equation}
\alpha^2\geq\frac{3}{2\sqrt{2}}\kappa\,{\tt M}\geq\frac{3}{4}\kappa
\inf\frac{\int_\Sigma t^aT_{ab}l^b\,{\rm d}\Sigma}{\int_\Sigma t_el^e
\,{\rm d}\Sigma} \label{eq:4.11}
\end{equation}
follows. In the special case of the vanishing extrinsic curvature 
the expression on the right is not less than Friedrich's sharp 
lower bound, and hence ${\tt M}$ also provides a sharp lower bound 
for the eigenvalues. 

However, now we show that $\frac{3}{2\sqrt{2}}\kappa\,{\tt M}$ is 
not only a lower bound, but it is just {\em the smallest eigenvalue} 
of $2{\cal D}^*{\cal D}$. In fact, since $2{\cal D}^*{\cal D}$ has a 
purely discrete spectrum with eigenvalues $\alpha^2$, the 
corresponding eigenspinors $\lambda^A_{\alpha^2}$ form a basis in $L_2
(\Sigma,\mathbb{S}^A)$. (The different eigenspinors corresponding to 
the same eigenvalue with higher multiplicity are chosen to be 
orthogonal to each other. See Appendix \ref{sub-A.4}.) Thus, if 
$\lambda^A$ is a smooth spinor field that satisfy (\ref{eq:3.6}) (or, 
equivalently, (\ref{eq:3.7})), then we can write $\lambda^A=\sum
_{\alpha^2\in[0,\infty)}c_{\alpha^2}\lambda^A_{\alpha^2}$ for some complex 
coefficients $c_{\alpha^2}$. Substituting this form of $\lambda^A$ 
into (\ref{eq:3.7}), we obtain 

\begin{equation*}
0=\langle\Delta^A{}_B\lambda^B,\lambda^A\rangle=\sum_{\alpha^2\in
[0,\infty)}\vert c_{\alpha^2}\vert^2\bigl(\alpha^2-\frac{3}{2\sqrt{2}}
\kappa\,{\tt M}\bigr)\Vert\lambda^A_{\alpha^2}\Vert^2_{L_2}.
\end{equation*}
Since by (\ref{eq:4.11}) $\alpha^2\geq\frac{3}{2\sqrt{2}}\kappa\,
{\tt M}$ holds for all eigenvalues $\alpha^2$, this implies that 
$\frac{3}{2\sqrt{2}}\kappa\,{\tt M}$ is just the smallest eigenvalue 
and $\lambda^A$ is a corresponding eigenvalue, otherwise $\lambda^A$ 
would have to be vanishing. 

The bound on the right hand side of (\ref{eq:4.11}) has been given 
in \cite{HiZh}, though the line of derivation was different. Here 
we show how the bound of \cite{HiZh} can be obtained in the present 
formalism. The key object is the one parameter family of differential 
operators 

\begin{equation}
\tilde{\cal D}_e\lambda^A:={\cal D}_e\lambda^A+sP^{AA'}_e{\cal D}
_{A'B}\lambda^B, \label{eq:4.12}
\end{equation}
labelled by the real parameter $s$. Then a direct calculation gives 

\begin{equation}
\Vert\tilde{\cal D}_e\lambda^A\Vert^2_{L_2}-2s(1+\frac{s}{4})
\Vert{\cal D}_{A'A}\lambda^A\Vert^2_{L_2}=\Vert{\cal D}_e\lambda^A
\Vert^2_{L_2}, \nonumber 
\end{equation}
by means of which from (\ref{eq:4.9}) we obtain 

\begin{equation}
(1+s+\frac{3}{4}s^2)\alpha^2\Vert\lambda^A\Vert^2_{L_2}=\Vert\tilde
{\cal D}_e\lambda^A\Vert^2_{L_2}+\frac{\kappa}{2\sqrt{2}}\int_\Sigma 
t^aT_{aBB'}\lambda^B\bar\lambda^{B'}{\rm d}\Sigma. \label{eq:4.13} 
\end{equation}
Thus the norm $\Vert\tilde{\cal D}_e\lambda^A\Vert^2_{L_2}$ as a 
function of the parameter $s$ on the right has a minimum precisely 
when $(1+s+\frac{3}{4}s^2)$ has, i.e. at $s=-\frac{2}{3}$. With 
this substitution (\ref{eq:4.13}) gives (\ref{eq:4.10}), and hence 
the bound (\ref{eq:4.11}). Indeed, the $L_2$-norm of $\tilde{\cal 
D}_e\lambda^A$ with the critical value of the parameter, $s=-
\frac{2}{3}$, is just the $L_2$-norm of the 3-surface twistor 
operator: $\Vert\tilde{\cal D}_e\lambda^A\Vert_{L_2}=\Vert{\cal D}
_{(AB}\lambda_{C)}\Vert_{L_2}$. Therefore, the role of the parameter 
$s$ in (\ref{eq:4.12}) is to change the Sen--Witten part in the 
$\langle\,\cdot,\cdot\,\rangle$--orthogonal decomposition 
(\ref{eq:3.1}) of the Sen derivative ${\cal D}_e\lambda^A$. For 
the critical value, $s=-\frac{2}{3}$, the `Sen--Witten content' 
of $\tilde{\cal D}_e$ is zero, i.e. the Sen--Witten operator built 
from $\tilde{\cal D}_e$ is vanishing: $\tilde{\cal D}_{A'A}\lambda
^A=0$ for any spinor field $\lambda^A$. 


\subsubsection{On the eigenvalue problem for the 3-surface 
twistor operator}
\label{sub-4.2.2}

The formal adjoint ${\cal T}^*:C^\infty(\Sigma,\mathbb{S}_{(ABC)})
\rightarrow C^\infty(\Sigma,\mathbb{S}_A)$ of the 3-surface twistor 
operator is ${\cal T}^*:\phi_{ABC}\mapsto{}^+{\cal D}^{BC}\phi_{ABC}$, 
where 

\begin{equation*}
{}^\pm{\cal D}_{AB}\lambda_C:=D_{AB}\lambda_C\mp\frac{1}{\sqrt2}
\chi_{ABC}{}^D\lambda_D,
\end{equation*}
the self-dual/anti-self-dual Sen connection, and ${\cal D}_{AB}={}
^-{\cal D}_{AB}$. Here $D_{AB}$ and $\chi_{ABCD}$ are the unitary 
spinor form of the intrinsic Levi-Civita derivative operator and 
the extrinsic curvature, respectively; and $C^\infty(\Sigma,
\mathbb{S}_{(ABC)})$ denotes the space of the totally symmetric 
three-index spinor fields on $\Sigma$. 

Repeating the calculations of subsection \ref{sub-2.2} for the 
3-surface twistor operator, it is easy to see that ${\cal T}^*{\cal 
T}$ is formally self-adjoint and positive. (It might be worth noting 
that the elliptic operator on which the analysis of \cite{BaKr} is 
based is just ${\cal T}^*{\cal T}$.) Moreover, by the definition of 
the formal adjoint, (\ref{eq:3.3}) can be rewritten in the form 

\begin{eqnarray}
\frac{4\sqrt{2}}{3\kappa}\Vert{\cal D}_{A'A}\lambda^A\Vert^2_{L_2}
 \!\!\!\!&=\!\!\!\!&\frac{\sqrt{2}}{\kappa}\langle{}^+{\cal D}^{BC}
 {\cal D}_{(AB}\lambda_{C)},\lambda_A\rangle+\int_\Sigma t^aT_{aBB'}
 \lambda^A\bar\lambda^{A'}{\rm d}\Sigma= \nonumber \\
 \!\!\!\!&=\!\!\!\!&\langle\frac{\sqrt{2}}{\kappa}{}^+{\cal D}^{BC}
 {\cal D}_{(AB}\lambda_{C)}+t_eT^{eBB'}G_{B'A}\lambda_B\, ,\,\lambda
 _A\rangle \geq {\tt M}\,\Vert\lambda^A\Vert^2_{L_2}. \label{eq:4.14}
\end{eqnarray}
The functional analytic properties of ${\cal T}^*{\cal T}$ are 
proven in the appendix, and are summarized in Appendix \ref{sub-A.4}: 
it is a positive, self-adjoint operator with a pure point spectrum. 
We define the eigenvalue problem for the 3-surface twistor operator by 
the convention $2{}^+{\cal D}^{BC}{\cal D}_{(AB}\lambda_{C)}=\tau^2\lambda
_A$, and hence {\em in vacuum} $\tau^2\geq\sqrt{2}\kappa \,{\tt M}$. 

However, equation (\ref{eq:4.14}) suggests the introduction of the 
operator ${\cal M}:{\rm Dom}({\cal T}^*{\cal T})\rightarrow L_2(
\Sigma,\mathbb{S}_A)$ defined by the expression in the first 
argument of the scalar product in the second line of (\ref{eq:4.14}). 
Clearly, this is the operator that is `behind' the lower bound ${\tt 
M}$, which is ${\cal T}^*{\cal T}$ `perturbed' by a bounded, 
positive zeroth order operator. Hence the key functional analytic 
properties of ${\cal M}$ are the same as those of ${\cal T}^*{\cal 
T}$. If we define the eigenvalue problem for ${\cal M}$ by ${\cal 
M}_A{}^B\lambda_B=\mu^2\lambda_A$, then by (\ref{eq:4.14}) and the 
definition of ${\tt M}$ 

\begin{equation*}
\frac{4\sqrt{2}}{3\kappa}\Vert{\cal D}_{A'A}\lambda^A\Vert^2_{L_2}=
\langle{\cal M}_A{}^B\lambda_B\, ,\,\lambda_A\rangle=\mu^2\Vert
\lambda^A\Vert^2_{L_2}\geq {\tt M}\,\Vert\lambda^A\Vert^2_{L_2},
\end{equation*}
and hence $\mu^2\geq{\tt M}$. Choosing $\lambda^A$ to be a spinor 
field that saturates the inequality on the right hand side (i.e. 
that satisfies (\ref{eq:3.7})), expanding it in terms of the 
eigenspinors of ${\cal M}$ and repeating the argumentation of the 
previous subsection, we find that ${\tt M}$ is just the smallest 
eigenvalue of ${\cal M}$ and $\lambda_A$ is a corresponding 
eigenspinor. 


\subsection{Examples}
\label{sub-4.3}

First, let $\Sigma$ be a $t={\rm const}$ spacelike hypersurface in 
a $k=1$ Friedmann--Robertson--Walker cosmological spacetime. It is 
homeomorphic to $S^3$, the intrinsic metric $h_{ab}$ is the standard 
3-sphere metric with scalar curvature $R={\rm const}$, and the 
extrinsic curvature is $\chi_{ab}=\frac{1}{3}\chi h_{ab}$ with $\chi
={\rm const}$. For this data set $t^a\,{}^4G_{ab}P^b_c=0$ and $-t^a
t^b\,{}^4G_{ab}=\frac{1}{2}R+\frac{1}{3}\chi^2={\rm const}$, and 
hence the lower bound on the right hand side of (\ref{eq:4.11}) is 
$\frac{3}{8}R+\frac{1}{4}\chi^2$. 
On the other hand, we know that this example with $\chi=0$ saturates 
the inequality of Friedrich, i.e. the smallest eigenvalue of the 
(Riemannian) eigenvalue problem $2D^{AA'}D_{A'B}\lambda^B=\beta^2
\lambda^A$ is just $\beta^2_1=\frac{3}{8}R$. We show that the 
corresponding eigenspinor is also an eigenspinor of $2{\cal D}^{AA'}
{\cal D}_{A'B}$, and the corresponding eigenvalue saturates both 
inequalities of (\ref{eq:4.11}). 
In fact, since $\chi={\rm const}$, $2{\cal D}^{AA'}{\cal D}_{A'B}
\lambda^B=2D^{AA'}D_{A'B}\lambda^B+\frac{1}{4}\chi^2\lambda^A$ holds, 
and hence for the smallest eigenvalue of $2{\cal D}^*{\cal D}$ we 
obtain $\alpha^2_1=\beta^2_1+\frac{1}{4}\chi^2$, just the lower bound 
on the right of (\ref{eq:4.11}). This shows, in particular, that the 
Witten equation does not have any non-trivial solution. The extrinsic 
curvature shifted both Friedrich's lower bound and the smallest 
Riemannian eigenvalue by the same positive term $\frac{1}{4}\chi^2$. 
It is easy to see that the 3-surface twistor operator annihilates 
this eigenspinor: since it is annihilated by the Riemannian 
3-surface twistor operator and ${\cal D}_{AB}\lambda_C=D_{AB}\lambda
_C+\frac{1}{6\sqrt{2}}\chi(2\varepsilon_{BC}\lambda_A+\varepsilon
_{AB}\lambda_C)$ holds, ${\cal D}_{(AB}\lambda_{C)}=0$ follows. Thus 
the first eigenvalue of $2{\cal D}^*{\cal D}$ coincides with 
$\frac{3}{2\sqrt{2}}\kappa\,{\tt M}$. The corresponding 
eigenspinor has constant components in the spin frame adapted to 
the globally defined left invariant orthonormal triad on $\Sigma$, 
and hence, in particular, it has no zeros. 

In the Bianchi I. cosmological model with toroidal spatial topology 
the solutions of $D_{A'A}\lambda^A=0$ are the spatially constant 
spinor fields. Clearly, these spinor fields solve $2{\cal D}^{AA'}
{\cal D}_{A'B}\lambda^B=\frac{1}{4}\chi^2\lambda^A$ too, i.e. the 
smallest eigenvalue of $2{\cal D}^*{\cal D}$ is $\frac{1}{4}\chi^2$. 
Thus, in particular, the Witten equation has a non-trivial solution 
precisely when $\chi=0$. Since by the Hamiltonian constraint 
(\ref{eq:2.9a}) we have that $\chi^2=2\kappa\mu+\chi_{ab}\chi^{ab}$, 
this, together with the dominant energy condition, implies $\chi
_{ab}=0$ and $\mu=0$, and hence the flatness of the spacetime. (For 
a review of the various Bianchi cosmological models, see e.g. 
\cite{Wald}.) 

The general results of section \ref{sec-4} and the specific 
properties of the eigenspinors discussed in these examples raise 
the possibility to generalize Witten's gauge condition, which, at 
the same time, is a modified version of Nester's gauge condition: 
Let the spinor field be the eigenspinor of the Sen--Witten 
operator with the {\em smallest non-negative eigenvalue}. 


\section{Appendix: An analysis of ${\cal D}$ and ${\cal T}$}
\label{sec-A}

In this appendix we recall some elementary properties of the 
Sen--Witten and 3-surface twistor operators (subsection 
\ref{sub-A.0}), quote the key Sobolev embedding theorems 
(subsection \ref{sub-A.1}) and we derive the estimates 
(subsection \ref{sub-A.2}) that we use to prove various 
properties of the operators ${\cal D}$, ${\cal T}$, ${\cal D}
^*{\cal D}$ and ${\cal T}^*{\cal T}$ (subsections \ref{sub-A.3} 
and \ref{sub-A.4}). In particular, since ${\cal T}$ is only 
overdetermined elliptic, the standard results of the theory of 
elliptic operators cannot be applied to it directly. However, as 
a consequence of the Sen--Witten identity, Einstein's equation 
and the dominant energy condition, we can derive a fundamental 
estimate for ${\cal T}$. As a consequence of this, we can prove 
that ${\cal T}$ shares most of the properties of the elliptic 
${\cal D}$. In some of these proofs we followed the logic of the 
proofs of some of the analogous statements for the Dirac operator 
acting on Dirac spinors given in \cite{FaSc}.


\subsection{Elementary analytic properties of ${\cal D}$ and 
${\cal T}$}
\label{sub-A.0}

The principal symbol of ${\cal D}$ is isomorphism, while the symbol 
of the 3-surface twistor operator ${\cal T}$ is only injective, but 
not surjective. Thus while the former is {\em elliptic}, the latter 
is only {\em overdetermined elliptic} (see e.g. \cite{Besse}, pp 462). 
By (\ref{eq:3.2}) both ${\cal D}:C^\infty(\Sigma,\mathbb{S}^A)
\rightarrow C^\infty(\Sigma,\bar{\mathbb{S}}_{A'})$ and ${\cal T}:C
^\infty(\Sigma,\mathbb{S}_A)\rightarrow C^\infty(\Sigma,\mathbb{S}
_{(ABC)})$ are bounded in the $H_1$-Sobolev norm, 
defined\footnote{Strictly speaking, the usual definition of the 
Sobolev norms (that we also adopt here) is dimensionally {\em not} 
consistent with the physical view that we can add quantities only 
with the {\em same} physical dimension. In fact, the dimensionally 
correct definition of the $H_k$--Sobolev norm would be $\Vert
\lambda^A\Vert_{L_2}+L\Vert{\cal D}_e\lambda^A\Vert_{L_2}+\cdots+
L^k\Vert{\cal D}_{e_1}\cdots{\cal D}_{e_k}\lambda^A\Vert_{L_2}$, 
where $L$ is a positive constant with ${length}$ physical dimension. 
Since in {\em classical} physics there is no such universal constant, 
the $H_k$--Sobolev norms for $k\geq1$ are {\em not canonically 
defined}. Therefore, it is only the $L_2$, but {\em not} the 
$H_k$--Sobolev norms for $k\geq1$, that can have physical meaning. 
Consequently, the operator norm of the (already) bounded operators 
${\cal D}$ and ${\cal T}$ coming from the $H_1$-norm does not seem 
to have physical meaning.} 
by $\Vert\lambda^A\Vert_{H_1}:=\Vert\lambda^A\Vert_{L_2}+\Vert{\cal 
D}_e\lambda^A\Vert_{L_2}$. Hence these operators can be extended in 
a unique way to be bounded linear operators ${\cal D}:H_1(\Sigma,
\mathbb{S}^A)\rightarrow L_2(\Sigma,\bar{\mathbb{S}}_{A'})$ and 
${\cal T}:H_1(\Sigma,\mathbb{S}_A)\rightarrow L_2(\Sigma,\mathbb{S}
_{(ABC)})$, respectively. Since ${\cal D}_e$ annihilates $\varepsilon
_{AB}$, it is natural to identify $H_1(\Sigma,\mathbb{S}^A)$ with 
$H_1(\Sigma,\mathbb{S}_A)$ via $\lambda^A\mapsto\lambda^B\varepsilon
_{BA}$. On the other hand, since ${\cal D}_{AA'}G_{BB'}=\sqrt{2}\chi
_{ab}$, it does {\em not} seem useful to identify the complex 
conjugate Sobolev space $H_1(\Sigma,\bar{\mathbb{S}}_{A'})$ with 
$H_1(\Sigma,\mathbb{S}^A)$ via $G_{AA'}$, and hence we keep them 
different. 

In subsection \ref{sub-2.2} we calculated the formal adjoint 
${\cal D}^*$ of ${\cal D}$, and we found that it is essentially 
($-1$ times) of the complex conjugate of ${\cal D}$ itself. Thus 
although ${\rm i}{\cal D}$ appears to be formally self-adjoint 
at first glance, as we saw in subsection \ref{sub-4.1.1}, as a 
consequence of ${\cal D}_{AA'}G_{BB'}=\sqrt{2}\chi_{ab}\not=0$, 
strictly speaking it is not even symmetric. The principal symbol 
of the formal adjoint ${\cal T}^*$ of the 3-surface twistor 
operator (determined in subsection \ref{sub-4.2.2}) is only 
surjective but not injective, and hence it is not elliptic 
either. It is only {\em underdetermined elliptic}.


\subsection{The Sobolev embedding theorems}
\label{sub-A.1}

The embedding theorems that we need state how the various function 
spaces over the same {\em compact} domain are related to each other. 
Since we need only the Sobolev spaces based on the $L_2$-norm, here 
we concentrate only on these special spaces. An extended discussion 
of these theorems and the related concepts can be found e.g. in 
\cite{Adams:Fournier,Evans}. 

\begin{theorem} Let $E(M)$ be a vector bundle over a compact, 
$m$-dimensional manifold $M$. Suppose that the space of its cross 
sections is endowed with a global Hermitian scalar product, we 
denote the $k$th Sobolev space of its cross sections by $H_k(M,E)$ 
and adopt the convention $H_0(M,E):=L_2(M,E)$. 

\begin{description}
\item[1.] Let $k\geq0$, $l>0$ be integers. Then the injection $i:H
   _{k+l}(M,E)\rightarrow H_k(M,E)$ is a dense, compact and continuous 
   embedding (i.e. for any $\phi\in H_k(M,E)$ and $\epsilon>0$ there 
   exists a cross section $\psi\in H_{k+l}(M,E)$ such that $\Vert\phi
   -\psi\Vert_{H_k}<\epsilon$; and if $\{\phi_i\}$, $i\in\mathbb{N}$, 
   is any sequence in $H_{k+l}(M,E)$ for which $\Vert\phi_i\Vert_{H
   _{k+l}}\leq1$, then there is a subsequence of $\{\phi_i\}$ which 
   is convergent in $H_k(M,E)$). 
\item[2.] The injection $i:H_{k+[\frac{m}{2}]+1}(M,E)\rightarrow C^k(M,
   E)$ is a dense, compact and continuous embedding, where 
   $[\frac{m}{2}]$ denotes the integer part of $\frac{1}{2}\dim M$, 
   and the norm on $C^k(M,E)$ is the $C^k$-supremum norm. 
\end{description}
\label{th:A.1.1}
\end{theorem}
\noindent
These statements are the Sobolev embedding theorems. The first is 
analogous to the inclusion of the space of the $C^{k+l}$ 
differentiable functions in the space of $C^k$ differentiable ones. 
By the second statement cross sections with $H_{k+[\frac{m}{2}]+1}$ 
control are $C^k$ cross sections in the classical sense. The 
compactness properties of the embeddings are known as the Rellich 
lemma.

\subsection{Elliptic estimates}
\label{sub-A.2}

We assume that the Einstein equations hold and that the matter 
fields satisfy the dominant energy condition. The next estimates 
are proven under these assumptions. 

The first of these, the so-called fundamental elliptic estimate, 
both for the Sen--Witten and the 3-surface twistor operators, are 
simple consequences of the definitions and equations (\ref{eq:3.2}) 
and (\ref{eq:3.3}): 

\begin{lemma} There is a positive constant $C$ such that for any 
  $\lambda^A\in H_1(\Sigma,\mathbb{S}^A)$ the inequalities 
\begin{equation}
 \Vert\lambda^A\Vert_{H_1}\leq\sqrt{2}\Vert{\cal D}_{A'A}\lambda^A
 \Vert_{L_2} +\Vert\lambda^A\Vert_{L_2}, \hskip 20pt
\Vert\lambda^A\Vert_{H_1}\leq\sqrt{\frac{3}{2}}\Vert{\cal D}_{(AB}
 \lambda_{C)}\Vert_{L_2} +C\Vert\lambda_A\Vert_{L_2}
\end{equation}
 hold. \hfill $\Box$
\label{l:A.2.1}
\end{lemma}
\noindent
Here the constant $C$ can be given explicitly: It is $C=1+\sqrt{
\frac{1}{2\sqrt{2}}\kappa T}$, where $T$ is a positive constant 
such that $t^bT_{bAA'}$, as a pointwise non-negative Hermitian 
scalar product on the spinor spaces, is not greater than $TG
_{AA'}$ everywhere on the {\em compact} $\Sigma$. Thus, although 
the 3-surface twistor operator is {\em not} elliptic, it is only 
{\em overdetermined elliptic}, by the Sen--Witten identity, 
Einstein's equation and the dominant energy condition we do have 
an estimate for it. For the sake of simplicity we call it a 
fundamental elliptic estimate, too. 

The second estimate, the so-called elliptic regularity estimate, 
is, in some sense, a generalization of the previous one. Again, 
though ${\cal T}$ is not elliptic, we have an estimate for that 
also: 

\begin{lemma} There are positive constants $C_1$, $C'_1$, $C_2$ 
  and $C'_2$ such that for any $\lambda^A\in H_k(\Sigma,\mathbb{S}
  ^A)$, $k\geq1$, for which ${\cal D}_{A'A}\lambda^A\in H_k(\Sigma,
  \bar{\mathbb{S}}_{A'})$ or ${\cal D}_{(AB}\lambda_{C)}\in H_k
  (\Sigma,\mathbb{S}_{(ABC)})$ is also true, the inequalities 
\begin{equation}
 \Vert\lambda^A\Vert_{H_{k+1}}\leq C_1\Vert{\cal D}_{A'A}\lambda^A
 \Vert_{H_k}+C'_1\Vert\lambda^A\Vert_{H_k}, \hskip 20pt
\Vert\lambda^A\Vert_{H_{k+1}}\leq C_2\Vert{\cal D}_{(AB}\lambda_{C)}
 \Vert_{H_k} +C'_2\Vert\lambda_A\Vert_{H_k},
\end{equation}
 respectively, hold.
\label{l:A.2.2}
\end{lemma}
{\it Proof:} Since $\Sigma$ is orientable and three dimensional, 
it is parallelizable, and hence we can find a {\em globally 
defined} $h_{ab}$-orthonormal dual frame field $\{e^a_{\bi},
\vartheta^{\bi}_a\}$, ${\bi}=1,2,3$, on $\Sigma$. Then by the 
triangle inequality 

\begin{align}
&\Vert{\cal D}_f{\cal D}_{e_1}\cdots{\cal D}_{e_k}\lambda^A\Vert
 _{L_2}=\Vert{\cal D}_f\bigl(\vartheta^{{\bi}_1}_{e_1}\cdots
 \vartheta^{{\bi}_k}_{e_k}e^{f_1}_{{\bi}_1}\cdots e^{f_k}_{{\bi}_k}{\cal 
 D}_{f_1}\cdots{\cal D}_{f_k}\lambda^A\bigr)\Vert_{L_2}\leq \nonumber \\
&\leq\Vert\bigl({\cal D}_f\vartheta^{\bi}_{e_1}\bigr)e^{f_1}_{\bi}
 {\cal D}_{f_1}{\cal D}_{e_2}\cdots{\cal D}_{e_k}\lambda^A\Vert_{L_2}
 +\cdots+\Vert\bigl({\cal D}_f\vartheta^{\bi}_{e_k}\bigr)e^{f_k}
 _{\bi}{\cal D}_{e_1}\cdots{\cal D}_{e_{k-1}}{\cal D}_{f_k}\lambda^A
 \Vert_{L_2}+ \nonumber \\
&+\Vert\vartheta^{{\bi}_1}_{e_1}\cdots\vartheta^{{\bi}_k}_{e_k}{\cal 
 D}_f\bigl(e^{f_1}_{{\bi}_1}\cdots e^{f_k}_{{\bi}_k}{\cal D}_{f_1}\cdots
 {\cal D}_{f_k}\lambda^A\bigr)\Vert_{L_2}. \label{eq:A.2.1}
\end{align}
Since $\Sigma$ is compact, there exists a constant $\tilde C_1>0$ 
such that $H^{\bi\bj}:=h^{ab}h^{cd}({\cal D}_a\vartheta_c^{\bi})({\cal 
D}_b\vartheta_d^{\bj})$, as a quadratic form, is nowhere greater 
than $\tilde C^2_1\delta^{\bi\bj}$ on $\Sigma$. Thus the first $k$ 
terms on the right of (\ref{eq:A.2.1}) can be estimated in this way 
to obtain 

\begin{eqnarray}
\Vert{\cal D}_f{\cal D}_{e_1}\cdots{\cal D}_{e_k}\lambda^A\Vert
 _{L_2}\!\!\!\!&\leq\!\!\!\!& k\tilde C_1\Vert{\cal D}_{e_1}\cdots
 {\cal D}_{e_k}\lambda^A\Vert_{L_2}+ \nonumber \\
\!\!\!\!&+\!\!\!\!&\Bigl(\sum_{{\bi}_1,...,{\bi}_k=1}^3\Vert{\cal 
 D}_c\bigl(e^{e_1}_{{\bi}_1}\cdots e^{e_k}_{{\bi}_k}{\cal D}_{e_1}
 \cdots{\cal D}_{e_k}\lambda^A\bigr)\Vert^2_{L_2}\Bigr)
 ^{\frac{1}{2}}, \label{eq:A.2.2}
\end{eqnarray}
where we rewrote the last term also using the definition of the $L
_2$-norm and the orthonormality of the dual frame field $\{e^a_{\bi},
\vartheta^{\bi}_a\}$. The next step is the use of the fundamental 
elliptic estimate in the last term on the right of (\ref{eq:A.2.2}). 
Thus, at this point, the detailed proof splits according to the 
two basic estimates, but the spirit of the proof in the two cases 
are the same. Here we present the detailed proof only in the case 
of the 3-surface twistor operator. 

Thus, in the last term of (\ref{eq:A.2.2}), let us use the estimate 

\begin{eqnarray*}
\Vert{\cal D}_e(e^{e_1}_{{\bi}_1}\cdots e^{e_k}_{{\bi}_k}{\cal 
D}_{e_1}\cdots{\cal D}_{e_k}\lambda_A)\Vert^2_{L_2}\!\!\!\!&\leq
\!\!\!\!&\frac{3}{2}\Vert{\cal D}_{(AB\vert}(e^{e_1}_{{\bi}_1}
\cdots e^{e_k}_{{\bi}_k}{\cal D}_{e_1}\cdots{\cal D}_{e_k}\lambda
_{\vert C)})\Vert^2_{L_2}+ \\
\!\!\!\!&\!\!\!\!&+\frac{1}{2\sqrt{2}}\kappa T\Vert e^{e_1}
_{{\bi}_1}\cdots e^{e_k}_{{\bi}_k}{\cal D}_{e_1}\cdots{\cal D}
_{e_k}\lambda^A\Vert^2_{L_2}
\end{eqnarray*}
coming from (\ref{eq:3.2}) and (\ref{eq:3.3}), and where $T$ is 
given in the text following Lemma \ref{l:A.2.1}. Substituting 
this estimate into the second term on the right of (\ref{eq:A.2.2}), 
using the triangle inequality again and the orthonormality of 
the frame field $\{e^a_{\bi}\}$ we obtain 

\begin{eqnarray}
\Vert{\cal D}_f{\cal D}_{e_1}\cdots{\cal D}_{e_k}\lambda^A\Vert
 _{L_2}\!\!\!\!&\leq\!\!\!\!&\bigl(k\tilde C_1+\sqrt{
 \frac{\kappa}{2\sqrt{2}}T}\bigr)\Vert{\cal D}_{e_1}\cdots
 {\cal D}_{e_k}\lambda^A\Vert_{L_2}+ \nonumber \\
\!\!\!\!&+\!\!\!\!&\frac{1}{\sqrt{6}}\Bigl(\sum_{{\bi}_1,...,
 {\bi}_k=1}^3\Vert3{\cal D}_{(AB\vert}(e^{e_1}_{{\bi}_1}\cdots e
 ^{e_k}_{{\bi}_k}{\cal D}_{e_1}\cdots{\cal D}_{e_k}\lambda_{\vert C)})
 \Vert^2_{L_2}\Bigr)^{\frac{1}{2}}. \label{eq:A.2.3}
\end{eqnarray}
The spinor field in the second term on the right is the totally 
symmetric part (in the spinor indices) of 

\begin{align}
&{\cal D}_{AB}\bigl(e^{e_1}_{{\bi}_1}\cdots e^{e_k}_{{\bi}_k}{\cal D}
 _{e_1}\cdots{\cal D}_{e_k}\lambda_C\bigr)=e^{e_1}_{{\bi}_1}\cdots e
 ^{e_k}_{{\bi}_k}\bigl({\cal D}_{AB}{\cal D}_{e_1}\cdots{\cal D}_{e_k}
 \lambda_C\bigr)+ \label{eq:A.2.4} \\
&+\bigl({\cal D}_{AB}e^{e_1}_{{\bi}_1}\bigr)e^{e_2}_{{\bi}_2}\cdots e
 ^{e_k}_{{\bi}_k}\bigl({\cal D}_{e_1}\cdots{\cal D}_{e_k}\lambda_C
 \bigr)+\cdots+\bigl({\cal D}_{AB}e^{e_k}_{{\bi}_k}\bigr)e^{e_1}
 _{{\bi}_1}\cdots e^{e_{k-1}}_{{\bi}_{k-1}}\bigl({\cal D}_{e_1}\cdots
 {\cal D}_{e_k}\lambda_C\bigr). 
 \nonumber 
\end{align}
The square of the $L_2$-norm of the second term on the right is 

\begin{eqnarray}
\Vert\bigl({\cal D}_{AB}e^{e_1}_{{\bi}_1}\bigr)e^{e_2}_{{\bi}_2}\cdots 
 e^{e_k}_{{\bi}_k}{\cal D}_{e_1}\cdots{\cal D}_{e_k}\lambda_C\Vert^2
 _{L_2}\!\!\!\!&= \!\!\!\!&
\int_\Sigma G^{AA'}G^{BB'}\bigl({\cal D}_{AB}e^{e_1}_{{\bi}_1}\bigr)
 \bigl({\cal D}_{A'B'}e^{f_1}_{{\bi}_1}\bigr)e^{e_2}_{{\bi}_2}e^{f_2}
 _{{\bi}_2}\cdots e^{e_k}_{{\bi}_k}e^{f_k}_{{\bi}_k} \times \nonumber \\
\!\!\!\!&\times\!\!\!\!& \bigl({\cal D}_{e_1}
 \cdots{\cal D}_{e_k}\lambda_C\bigr)\bigl({\cal D}_{f_1}\cdots{\cal 
 D}_{f_k}\bar\lambda_{C'}\bigr)G^{CC'}{\rm d}\Sigma. \nonumber
\end{eqnarray}
However, there exists a positive constant $\tilde C_2$ such that 
$G^{AA'}G^{BB'}({\cal D}_{AB}e^{e_1}_{{\bi}_1})({\cal D}_{A'B'}e^{f_1}
_{{\bi}_1})e^{e_2}_{{\bi}_2}e^{f_2}_{{\bi}_2}$ $\cdots e^{e_k}_{{\bi}_k}e
^{f_k}_{{\bi}_k}$, as a `multi-quadratic form', is not less than 
$\tilde C^2_2(-h^{e_1f_1})\cdots(-h^{e_kf_k})$ on $\Sigma$. With 
this bound we have that 

\begin{equation*}
\Vert\bigl({\cal D}_{AB}e^{e_1}_{{\bi}_1}\bigr)e^{e_2}_{{\bi}_2}\cdots 
e^{e_k}_{{\bi}_k}{\cal D}_{e_1}\cdots{\cal D}_{e_k}\lambda_C\Vert_{L_2}
\leq\tilde C_2\Vert{\cal D}_{e_1}\cdots{\cal D}_{e_k}\lambda^A\Vert
_{L_2}.
\end{equation*}
Clearly, we have similar estimates for the last $k-1$ terms on 
the right hand side of (\ref{eq:A.2.4}) also. Substituting all 
these into (\ref{eq:A.2.3}) and using the triangle inequality we 
obtain 

\begin{eqnarray}
\Vert{\cal D}_f{\cal D}_{e_1}\cdots{\cal D}_{e_k}\lambda^A\Vert
 _{L_2}\!\!\!\!&\leq\!\!\!\!&\bigl(k\tilde C_1+\sqrt{
 \frac{\kappa}{2\sqrt{2}}T}+\sqrt{\frac{3}{2}}k3^k\tilde C_2
 \bigr)\Vert{\cal D}_{e_1}\cdots{\cal D}_{e_k}\lambda^A\Vert_{L_2}
 + \nonumber \\
\!\!\!\!&+\!\!\!\!&\sqrt{\frac{3}{2}}\sum_{{\bi}_1,...,{\bi}_k=1}^3
 \Vert e^{e_1}_{{\bi}_1}\cdots e^{e_k}_{{\bi}_k}{\cal D}_{(AB\vert}
 {\cal D}_{e_1}\cdots{\cal D}_{e_k}\lambda_{\vert C)})\Vert_{L_2}. 
\nonumber
\end{eqnarray}
Estimating $e^{e_1}_{{\bi}_1}e^{f_1}_{{\bi}_1}\cdots e^{e_k}_{{\bi}_k}
e^{f_k}_{{\bi}_k}$ on $\Sigma$ by $\tilde C^2_3(-h^{e_1f_1})\cdots
(-h^{e_kf_k})$ (for some positive constant $\tilde C_3$) in the 
last term, we find 

\begin{equation}
\Vert{\cal D}_f{\cal D}_{e_1}\cdots{\cal D}_{e_k}\lambda^A\Vert
_{L_2}\leq\tilde C\Vert{\cal D}_{e_1}\cdots{\cal D}_{e_k}\lambda
^A\Vert_{L_2}+C_2\Vert{\cal D}_{(AB\vert}{\cal D}_{e_1}\cdots
{\cal D}_{e_k}\lambda_{\vert C)}\Vert_{L_2}, \label{eq:A.2.5}
\end{equation}
where we have relabelled the constants. 

Finally, let us write the spinor field in the last term of the 
estimate (\ref{eq:A.2.5}) as 

\begin{equation*}
{\cal D}_{AB}{\cal D}_{e_1}\cdots{\cal D}_{e_k}\lambda_C=\Bigl(
{\cal D}_{AB}{\cal D}_{e_1}-{\cal D}_{e_1}{\cal D}_{AB}\Bigr)
\bigl({\cal D}_{e_2}\cdots{\cal D}_{e_k}\lambda_C\bigr)+{\cal D}
_{e_1}\Bigl({\cal D}_{AB}{\cal D}_{e_2}\cdots{\cal D}_{e_k}
\lambda_C\Bigr),
\end{equation*}
and, using (\ref{eq:2.5}), express the commutator in terms of 
the curvature $F^A{}_{Bcd}$ and the extrinsic curvature $\chi
_{ab}$. Then by repeating this substitution, in finite steps we 
obtain that ${\cal D}_{(AB\vert}{\cal D}_{e_1}\cdots{\cal D}_{e_k}
\lambda_{\vert C)}$ is the sum of ${\cal D}_{e_1}\cdots{\cal D}
_{e_k}({\cal D}_{(AB}\lambda_{C)})$ and terms which contain at 
most the $k$th order derivative of $\lambda^A$, and the 
coefficients of the latter are built from the curvature, the 
extrinsic curvature and their (at most $(k-1)$st order) 
derivatives. Thus by the compactness of $\Sigma$ and the 
appropriate smoothness of the geometry there exist positive 
constants $\tilde{\tilde C}_0$, $\tilde{\tilde C}_1$, ..., 
$\tilde{\tilde C}_k$ such that 

\begin{eqnarray*}
\Vert{\cal D}_{(AB\vert}{\cal D}_{e_1}\cdots{\cal D}_{e_k}\lambda
 _{\vert C)}\Vert_{L_2}\!\!\!\!&\leq\!\!\!\!&\tilde{\tilde C}_0
 \Vert\lambda^A\Vert_{L_2}+\tilde{\tilde C}_1\Vert{\cal D}_e
 \lambda^A\Vert_{L_2}+\cdots+\tilde{\tilde C}_k\Vert{\cal D}
 _{e_1}\cdots{\cal D}_{e_k}\lambda^A\Vert_{L_2}+ \\
\!\!\!\!&+\!\!\!\!&\Vert{\cal D}_{e_1}\cdots{\cal D}_{e_k}{\cal 
 D}_{(AB}\lambda_{C)}\Vert_{L_2};
\end{eqnarray*}
which, together with (\ref{eq:A.2.5}) and the definition of the 
$H_k$-Sobolev norm, yield the desired estimate. \hfill $\Box$

\medskip

The $H_1$-norm of the spinor fields not belonging to the kernel 
of ${\cal D}$ or ${\cal T}$ can also be estimated by the $L
_2$-norm of these operators acting on the spinor fields.  This 
lemma is the adaptation of a result of \cite{FaSc}, given for 
the Riemannian Dirac operator acting on Dirac spinors, to the 
Sen--Witten and 3-surface twistor operators acting on Weyl 
spinors: 

\begin{lemma} There are positive constants $C_0$, and $C'_0$ such 
 that the inequalities 
\begin{align}
 &\Vert\lambda^A\Vert_{H_1}\leq C_0\Vert{\cal D}_{A'A}\lambda^A
 \Vert_{L_2}, \hskip 20pt \forall\lambda^A\in\ker({\cal D})^\bot
 \cap H_1(\Sigma,\mathbb{S}^A), \label{eq:A.2.6.a}\\
 &\Vert\lambda^A\Vert_{H_1}\leq C'_0\Vert{\cal D}_{(AB}\lambda
 _{C)}\Vert_{L_2}, \hskip 20pt \forall\lambda_A\in\ker({\cal T})
 ^\bot\cap H_1(\Sigma,\mathbb{S}_A) \label{eq:A.2.6.b}
\end{align}
 hold.
\label{l:A.2.3}
\end{lemma}
{\it Proof:} The proof for $(\ker({\cal D}))^\bot$ and $(\ker({\cal 
T}))^\bot$ are essentially the same, thus in the present proof 
$(\ker)^\bot$ may be any of them. 

Let us define $S:=\{\hat\lambda^A\in(\ker)^\bot\cap H_1(\Sigma,
\mathbb{S}^A)\,\vert\,\Vert\hat\lambda^A\Vert_{L_2}=1\,\}$, and the 
function $F:H_1\big(\Sigma,\mathbb{S}^A\big)\rightarrow[0,\infty)$ 
to be $\Vert{\cal D}_{A'A}\lambda^A\Vert_{L_2}$ or $\Vert{\cal D}
_{(AB}\lambda_{C)}\Vert_{L_2}$ in the two cases, respectively. Since 
${\cal D}$, ${\cal T}$ and the norm $\Vert.\Vert_{L_2}$ are continuous, 
the function $F$ is also continuous. First we show that the restriction 
of this function to $S$ has a {\em strictly positive} minimum, and 
there is a spinor field $\hat\phi^A\in S$ where $F$ takes this minimum 
value. 

Thus, let $F_0:=\lim\inf\{\,F(\hat\lambda^A)\,\vert\,\hat\lambda^A
\in S\,\}\geq0$, and let $\{\hat\phi^A_i\}$, $i\in\mathbb{N}$, be 
a sequence in $S$ such that $\{F(\hat\phi^A_i)\}$ is monotonically 
decreasing and $\lim_{i\rightarrow\infty}F(\hat\phi^A_i)=F_0$. Since 
$\{F(\hat\phi^A_i)\}$ is monotonically decreasing, it is bounded: 
$\exists K>0$ such that $F(\hat\phi^A_i)\leq K$ for all $i\in
\mathbb{N}$. Thus, by the fundamental elliptic estimates for 
${\cal D}$ and ${\cal T}$ (Lemma \ref{l:A.2.1}), 

\begin{align*}
&\Vert\hat\phi^A_i\Vert_{H_1}\leq\sqrt{2}\Vert{\cal D}_{A'A}\hat\phi
  ^A_i\Vert_{L_2}+\Vert\hat\phi^A_i\Vert_{L_2}\leq\sqrt{2}K+1, \\
&\Vert\hat\phi^A_i\Vert_{H_1}\leq\sqrt{\frac{3}{2}}\Vert{\cal D}
 _{(AB}\hat\phi_{\vert i\vert C)}\Vert_{L_2}+C\Vert\hat\phi^A_i
 \Vert_{L_2}\leq\sqrt{\frac{3}{2}}K+C,
\end{align*}
where $C$ is the positive constant of Lemma \ref{l:A.2.1}; i.e. 
$\{\hat\phi^A_i\}$ is a bounded sequence in $H_1(\Sigma,\mathbb{S}
^A)$. Thus there is a subsequence $\{\hat\phi^A_{i_k}\}$, $k\in
\mathbb{N}$, of $\{\hat\phi^A_i\}$ which converges in the weak 
topology of $H_1(\Sigma,\mathbb{S}^A)$ to some $\hat\phi^A_w\in H
_1(\Sigma,\mathbb{S}^A)$. 

On the other hand, by the Rellich lemma the injection $H_1(\Sigma,
\mathbb{S}^A)\rightarrow L_2(\Sigma,\mathbb{S}^A)$ is compact, and 
hence by the boundedness of the sequence $\{\hat\phi^A_{i_k}\}$ in 
the $H_1$-norm there is a subsequence $\{\hat\phi^A_{i_{k_l}}\}$, 
$l\in\mathbb{N}$, which converges in the $L_2$-norm to some $\hat
\phi^A_s\in L_2(\Sigma,\mathbb{S}^A)$. However, since the strong 
and the weak limits must be the same, we obtain that $\{\hat\phi
^A_{i_{k_l}}\}$ converges to some $\hat\phi^A:=\hat\phi^A_s=\hat\phi
^A_w\in H_1(\Sigma,\mathbb{S}^A)$ in the $L_2$-norm. Moreover, 
since the norm is continuous, $\Vert\hat\phi^A\Vert_{L_2}=\lim
_{l\rightarrow\infty}\Vert\hat\phi^A_{i_{k_l}}\Vert_{L_2}=1$. Also, since 
$\hat\phi^A_i\in(\ker)^\bot\cap H_1(\Sigma,\mathbb{S}^A)$ and $\hat
\phi^A_{i_{k_l}}\rightarrow\hat\phi^A$, the spinor field $\hat\phi^A$ 
is orthogonal to $\ker$. Thus, $\hat\phi^A\in S$. Finally, by 
the continuity of $F$, $F(\hat\phi^A)=\lim_{l\rightarrow\infty}F(\hat
\phi^A_{i_{k_l}})=F_0$. However, by $\Vert\hat\phi^A\Vert_{L_2}=1$ and 
$\hat\phi^A\in(\ker)^\bot$, this cannot be zero, i.e. $F_0>0$. 

Therefore, there exist positive constants $F_0$ and $F'_0$ such that 
$0<F_0<\Vert{\cal D}_{A'A}\hat\lambda^A\Vert_{L_2}$ and $0<F'_0<\Vert
{\cal D}_{(AB}\hat\lambda_{C)}\Vert_{L_2}$ for any $\hat\lambda^A\in S$; 
i.e. for any spinor field $\lambda^A\in(\ker({\cal D}))^\bot\cap H_1
(\Sigma,\mathbb{S}^A)$ or $\lambda_A\in(\ker({\cal T}))^\bot\cap H_1
(\Sigma,\mathbb{S}_A)$, respectively, one has 

\begin{equation*}
\Vert\lambda^A\Vert_{L_2}\leq\frac{1}{F_0}\Vert{\cal D}_{A'A}\lambda
^A\Vert_{L_2}, \hskip 20pt 
\Vert\lambda_A\Vert_{L_2}\leq\frac{1}{F'_0}\Vert{\cal D}_{(AB}\lambda
_{C)}\Vert_{L_2}.
\end{equation*}
Combining these with the corresponding fundamental elliptic 
estimates we obtain the estimates of the Lemma, where $C_0=\sqrt{2}
+(1/F_0)$ and $C'_0=\sqrt{\frac{3}{2}}+(C/F'_0)$. \hfill $\Box$


\subsection{Kernels and ranges, domains and Fredholm 
properties}
\label{sub-A.3}

First we prove statements on the structure of the kernel and 
range of the Sen--Witten and 3-surface twistor operators. 

\begin{proposition}
$\ker({\cal D})$ and $\ker({\cal T})$ are finite dimensional 
and their elements are smooth spinor fields.
\label{p:A.3.1}
\end{proposition}
{\it Proof:} We prove only $\dim\ker({\cal T})<\infty$, the 
proof of $\dim\ker({\cal D})<\infty$ is similar. Suppose, on 
the contrary, that $\ker({\cal T})$ is infinite dimensional, 
and let $\{\lambda^i_A\}$, $i\in\mathbb{N}$, be a sequence in 
$\ker({\cal T})$ such that $\langle\lambda^i_A,\lambda^j_A
\rangle=\delta^{ij}$, i.e. for example an $L_2$-orthonormal 
basis in $\ker({\cal T})$. Then by the fundamental elliptic 
estimate (Lemma \ref{l:A.2.1}) $\Vert\lambda^i_A\Vert_{H_1}
\leq C\Vert\lambda^i_A\Vert_{L_2}=C$, i.e. in particular this 
sequence is bounded in $H_1(\Sigma,\mathbb{S}_A)$. Since by 
the Rellich lemma (Theorem \ref{th:A.1.1}.1) the injection 
$H_1(\Sigma,\mathbb{S}_A)\rightarrow L_2(\Sigma,\mathbb{S}_A)$ 
is compact, there is a subsequence $\{\lambda^{i_k}_A\}$, $k\in
\mathbb{N}$, of $\{\lambda^i_A\}$ which is convergent in $L_2
(\Sigma,\mathbb{S}_A)$. Hence, it would have to be Cauchy. 
Since, however, $\Vert\lambda^i_A-\lambda^j_A\Vert^2_{L_2}=\langle
\lambda^i_A-\lambda^j_A,\lambda^i_A-\lambda^j_A\rangle=2$ holds 
for $i\not=j$, no subsequence of $\{\lambda^i_A\}$ could be 
Cauchy. Therefore, $\dim\ker({\cal T})<\infty$. 

To prove smoothness, suppose that $\lambda_A\in\ker({\cal T})
\subset H_1(\Sigma,\mathbb{S}_A)$. Then the elliptic regularity 
estimate (Lemma \ref{l:A.2.2}) yields that $\lambda_A\in H_2(
\Sigma,\mathbb{S}_A)$. This yields that $\lambda_A\in H_3(\Sigma,
\mathbb{S}_A)$, ... etc., i.e. that $\lambda_A\in H_k(\Sigma,
\mathbb{S}_A)$ for any $k\in\mathbb{N}$. Then by the Sobolev 
lemma (Theorem \ref{th:A.1.1}.2) this yields that $\lambda_A\in 
C^\infty(\Sigma,\mathbb{S}_A)$.  \hfill $\Box$

\medskip

\begin{proposition}
${\rm Im}({\cal D})\subset L_2(\Sigma,\bar{\mathbb{S}}_{A'})$ and 
${\rm Im}({\cal T})\subset L_2(\Sigma,\mathbb{S}_{(ABC)})$ are 
closed subspaces.
\label{p:A.3.2}
\end{proposition}
{\it Proof:} Let $\{\bar\phi^i_{A'}\}$ and $\{\phi^i_{ABC}\}$, $i\in
\mathbb{N}$, be Cauchy sequences in ${\rm Im}({\cal D})$ and ${\rm 
Im}({\cal T})$, respectively. Thus we may assume that 

\begin{equation*}
\bar\phi^i_{A'}={\cal D}_{A'A}\lambda^A_i, \hskip 20pt
\phi^i_{ABC}={\cal D}_{(AB}\mu^i_{C)},
\end{equation*}
where $\lambda^A_i\in(\ker({\cal D}))^\bot\cap H_1(\Sigma,\mathbb{S}
^A)$ and $\mu^i_A\in(\ker({\cal T}))^\bot\cap H_1(\Sigma,\mathbb{S}
_A)$. Then by Lemma \ref{l:A.2.3}

\begin{align*}
&\Vert\lambda^A_i-\lambda^A_j\Vert_{H_1}\leq C_0\Vert{\cal D}_{A'A}
 \lambda^A_i-{\cal D}_{A'A}\lambda^A_j\Vert_{L_2}=C_0\Vert\bar\phi
 ^i_{A'}-\bar\phi^j_{A'}\Vert_{L_2}, \\
&\Vert\mu^i_A-\mu^j_A\Vert_{H_1}\leq C'_0\Vert{\cal D}_{(AB}\mu^i
 _{C)}-{\cal D}_{(AB}\mu^j_{C)}\Vert_{L_2}=C'_0\Vert\phi^i_{ABC}-
 \phi^j_{ABC}\Vert_{L_2};
\end{align*}
and hence $\{\lambda^A_i\}$ and $\{\mu^i_A\}$ are Cauchy sequences 
in $H_1(\Sigma,\mathbb{S}^A)$. Thus they converge strongly to some 
$\lambda^A\in H_1(\Sigma,\mathbb{S}^A)$ and $\mu_A\in H_1(\Sigma,
\mathbb{S}_A)$, respectively. Therefore, since ${\cal D}$ and 
${\cal T}$ are continuous, $\bar\phi^i_{A'}={\cal D}_{A'A}\lambda^A
_i\rightarrow{\cal D}_{A'A}\lambda^A\in{\rm Im}({\cal D})$ and $\phi
^i_{ABC}={\cal D}_{(AB}\mu^i_{C)}\rightarrow{\cal D}_{(AB}\mu_{C)}\in
{\rm Im}({\cal T})$ when $i\rightarrow\infty$, i.e. ${\rm Im}(
{\cal D})$ and ${\rm Im}({\cal T})$ are closed. \hfill $\Box$

\medskip
By Proposition \ref{p:A.3.2} and Lemma \ref{l:A.2.3} it is easy to 
show that ${\cal D}$ and ${\cal T}$, as densely defined operators 
from $L_2(\Sigma,\mathbb{S}^A)$ to $L_2(\Sigma,\bar{\mathbb{S}}
_{A'})$ and to $L_2(\Sigma,\mathbb{S}_{(ABC)})$, respectively, are 
{\em closed operators}. (Recall that a linear operator $T:{\rm Dom}
(T)\subset X\rightarrow Y$ from the Banach space $X$ to the Banach 
space $Y$ is called {\em closed} if for every Cauchy sequence $x_i
\in{\rm Dom}(T)$ for which $Tx_i$ is also convergent (with the limit 
points $x\in X$ and $y\in Y$, respectively), $x\in{\rm Dom}(T)$ and 
$Tx=y$ follow. This is also equivalent to the statement that the 
graph, $G(T):=\{(x,Tx)\,\vert \, x\in{\rm Dom}(T)\,\}$, of $T$ is a 
closed subspace of $X\times Y$. See \cite{Ka} pp. 164, and in 
particular Problem 5.15 on page 165.) 

Since the Sen--Witten operator with the extended domain, ${\cal 
D}:H_1(\Sigma,\mathbb{S}^A)\rightarrow L_2(\Sigma,\bar{\mathbb{S}}
_{A'})$, is bounded with respect to the $H_1(\Sigma,\mathbb{S}^A)$ 
and $L_2(\Sigma,\bar{\mathbb{S}}_{A'})$ norms, at first sight it 
seems natural to consider its adjoint to be the uniquely determined 
bounded dual operator from $L_2(\Sigma,\bar{\mathbb{S}}_{A'})$ to 
$H_1(\Sigma,\mathbb{S}^A)$. However, this notion of adjoint would 
depend on the $H_1$-Sobolev norm, which, as we noted, does not have 
a well defined physical meaning. Moreover, this would not be an 
extension of the formal adjoint ${\cal D}^*$ given explicitly on 
the smooth spinor fields by $\bar\mu_{A'}\mapsto{\cal D}^{AA'}\bar
\mu_{A'}$. Thus, we do not follow this strategy. 

Recall that the formal adjoint was introduced by using only the 
(physically meaningful) $L_2$-norms, with respect to which ${\cal 
D}$ is {\em not} bounded. Thus ${\cal D}$ can be extended only to 
a proper dense subspace of $L_2(\Sigma,\mathbb{S}^A)$, but not to 
the whole of $L_2(\Sigma,\mathbb{S}^A)$. Therefore, though we 
still consider ${\cal D}$ to be extended to be a map $H_1(\Sigma,
\mathbb{S}^A)\rightarrow L_2(\Sigma,\bar{\mathbb{S}}_{A'})$, but 
from the point of view of its adjoint we consider $H_1(\Sigma,
\mathbb{S}^A)$ only to be a dense subspace of $L_2(\Sigma,
\mathbb{S}^A)$ and we do not use its $H_1$-Sobolev norm. Then, 
according to functional analysis (see e.g. \cite{Ka}, pp 167), 
the domain of the adjoint ${\cal D}^*$ of ${\cal D}:H_1(\Sigma,
\mathbb{S}^A)\rightarrow L_2(\Sigma,\bar{\mathbb{S}}_{A'})$ is 
defined by 

\begin{eqnarray}
{\rm Dom}\bigl({\cal D}^*\bigr):=\{\bar\mu_{A'}\in L_2(\Sigma,\bar{
 \mathbb{S}}_{A'})\,\vert\,\!\!\!\!&{}\!\!\!\!&\exists\,\nu^A\in L_2
 (\Sigma,\mathbb{S}^A): \label{eq:A.3.1} \\
\!\!\!\!&{}\!\!\!\!&\langle{\cal D}_{A'A}\lambda^A,\bar\mu_{A'}
 \rangle=\langle\lambda^A,\nu^A\rangle\,\forall\,\lambda^A\in H_1(
 \Sigma,\mathbb{S}^A)\,\}. \nonumber
\end{eqnarray}
Here the spinor field $\nu^A$ is uniquely determined and is 
necessarily orthogonal to $\ker({\cal D})$, and the adjoint 
operator ${\cal D}^*$ is defined to be the map $\bar\mu_{A'}\mapsto
\nu^A$. The following statement justifies this choice for the domain 
of ${\cal D}$ and the notion of the adjoint: Although ${\cal D}$ is 
{\em not} a formally self-adjoint operator on $C^\infty(\Sigma,
\mathbb{S}^A)$ (since e.g. the domain and range spaces consist of 
the cross sections of different vector bundles), the complex 
conjugate of its extended domain, i.e. $H_1(\Sigma,\bar{\mathbb{S}}
_{A'})$, is just the domain of the adjoint ${\cal D}^*$: 

\begin{proposition}
The domain of the adjoint ${\cal D}^*:{\rm Dom}({\cal D}^*)
\rightarrow L_2(\Sigma,\mathbb{S}^A)$ of ${\cal D}$ is just the 
complex conjugate of the domain of ${\cal D}$, i.e. ${\rm Dom}({\cal 
D}^*)=H_1(\Sigma,\bar{\mathbb{S}}_{A'})$. 
\label{p:A.3.3}
\end{proposition}
{\it Proof:} First recall that, for any $\lambda^A,\mu^A\in H_1(
\Sigma,\mathbb{S}^A)$, $\langle{\cal D}_{A'A}\lambda^A,\bar\mu_{A'}
\rangle=\langle\lambda^A,{\cal D}^{AA'}\bar\mu_{A'}\rangle$ holds. 
Thus, if $\bar\mu_{A'}\in H_1(\Sigma,\bar{\mathbb{S}}_{A'})$, then 
${\cal D}^{AA'}\bar\mu_{A'}\in L_2(\Sigma,\mathbb{S}^A)$, and hence 
with the notation $\nu^A:={\cal D}^{AA'}\bar\mu_{A'}$ one has that 
$\langle{\cal D}_{A'A}\lambda^A,\bar\mu_{A'}\rangle=\langle\lambda
^A,\nu^A\rangle$ for any $\lambda^A\in H_1(\Sigma,\mathbb{S}^A)$; 
i.e. by the definition (\ref{eq:A.3.1}) of ${\rm Dom}({\cal D}^*)$, 
$\bar\mu_{A'}\in {\rm Dom}({\cal D}^*)$. Therefore, $H_1(\Sigma,
\bar{\mathbb{S}}_{A'})\subset{\rm Dom}({\cal D}^*)$. 

Conversely, let $\bar\mu_{A'}\in{\rm Dom}({\cal D}^*)$. Since $H_1
(\Sigma,\bar{\mathbb{S}}_{A'})\subset L_2(\Sigma,\bar{\mathbb{S}}
_{A'})$ is dense, there exists a sequence $\{\bar\mu^i_{A'}\}$ in 
${\rm Dom}({\cal D}^*)\cap H_1(\Sigma,\bar{\mathbb{S}}_{A'})$, 
$i\in\mathbb{N}$, such that $\bar\mu^i_{A'}\rightarrow\bar\mu_{A'}$ 
in the $L_2$-norm as $i\rightarrow\infty$; moreover 

$$
\langle\lambda^A,{\cal D}^{AA'}\bar\mu^i_{A'}\rangle=\langle{\cal 
D}_{A'A}\lambda^A,\bar\mu^i_{A'}\rangle\rightarrow\langle{\cal D}
_{A'A}\lambda^A,\bar\mu_{A'}\rangle 
\hskip 20pt {\rm if}\,\,i\rightarrow\infty.
$$
By $\bar\mu_{A'}\in{\rm Dom}({\cal D}^*)$ and the definition of 
${\rm Dom}({\cal D}^*)$ there exists $\nu^A\in L_2(\Sigma,
\mathbb{S}^A)$ such that the limit on the right hand side has the 
form $\langle\lambda^A,\nu^A\rangle$; i.e. $\langle\lambda^A,{\cal 
D}^{AA'}\bar\mu^i_{A'}-\nu^A\rangle\rightarrow0$ for any $\lambda^A
\in H_1(\Sigma,\mathbb{S}^A)$ if $i\rightarrow\infty$. But since 
$H_1(\Sigma,\mathbb{S}^A)\subset L_2(\Sigma,\mathbb{S}^A)$ is 
dense, this also implies that $\langle\omega^A,{\cal D}^{AA'}\bar
\mu^i_{A'}-\nu^A\rangle\rightarrow0$ for any $\omega^A\in L_2(\Sigma,
\mathbb{S}^A)$ if $i\rightarrow\infty$; i.e. ${\cal D}^{AA'}
\bar\mu^i_{A'}\rightarrow\nu^A$ in the {\em weak topology} of $L_2
(\Sigma,\mathbb{S}^A)$. 

Since every weakly convergent sequence is bounded, there exist 
positive constants $K_1$ and $K_2$ such that $\Vert\bar\mu^i_{A'}
\Vert_{L_2}\leq K_1$ and $\Vert{\cal D}^{AA'}\bar\mu^i_{A'}\Vert
_{L_2}\leq K_2$ for any $i\in\mathbb{N}$. By the fundamental 
elliptic estimate for the Sen--Witten operator applied to the 
complex conjugate spinors these imply that the sequence $\{\bar
\mu^i_{A'}\}$ is bounded in the $H_1$-Sobolev norm as well. Thus, 
the sequence $\{\bar\mu^i_{A'}\}$ contains a subsequence $\{\bar
\mu^{i_k}_{A'}\}$, $k\in\mathbb{N}$, which converges weakly in 
$H_1(\Sigma,\bar{\mathbb{S}}_{A'})$ to some $\bar\mu^w_{A'}\in 
H_1(\Sigma,\bar{\mathbb{S}}_{A'})$. Since, however, the sequence 
$\{\bar\mu^i_{A'}\}$ was assumed to converge strongly to $\bar
\mu_{A'}$, the strong and the weak limits must coincide. Thus we 
obtained that $\bar\mu_{A'}=\bar\mu^w_{A'}\in H_1(\Sigma,\bar{
\mathbb{S}}_{A'})$, i.e. that ${\rm Dom}({\cal D}^*)\subset H_1
(\Sigma,\bar{\mathbb{S}}_{A'})$.   \hfill $\Box$

\medskip

In many applications the Fredholm property of operators plays a 
key role. (For a very readable review of the theory of Fredholm 
operators, see e.g. \cite{Pa}, Ch. VI.-VII.) Although the Fredholm 
property of ${\cal D}$ follows from the general elliptic theory, 
here we give a simple, direct proof.

\begin{proposition}
${\cal D}:H_1(\Sigma,\mathbb{S}^A)\rightarrow L_2(\Sigma,\bar{
\mathbb{S}}_{A'})$ is a Fredholm operator with zero analytic 
index. 
\label{p:A.3.4}
\end{proposition}
{\it Proof:} By Proposition \ref{p:A.3.1} $\ker({\cal D})$ is 
finite dimensional and by Proposition \ref{p:A.3.2} ${\rm Im}
({\cal D})$ is closed. Thus we need to show only that $\coker
({\cal D}):=L_2(\Sigma,\bar{\mathbb{S}}_{A'})/{\rm Im}({\cal 
D})$ is finite dimensional. 

A spinor field $\lambda^A$ belongs to $\ker({\cal D})$ precisely 
when $\langle{\cal D}_{A'A}\lambda^A,\bar\mu_{A'}\rangle=0$ for 
all $\bar\mu_{A'}\in L_2(\Sigma,\bar{\mathbb{S}}_{A'})$. Since, 
however, $H_1(\Sigma,\bar{\mathbb{S}}_{A'})\subset L_2(\Sigma,
\bar{\mathbb{S}}_{A'})$ is dense, $\lambda^A\in\ker({\cal D})$ is 
equivalent even to $0=\langle{\cal D}_{A'A}\lambda^A,\bar\phi_{A'}
\rangle=\langle\lambda^A,{\cal D}^{AA'}\bar\phi_{A'}\rangle$ for 
all $\bar\phi_{A'}\in H_1(\Sigma,\bar{\mathbb{S}}_{A'})$. Since 
by Proposition \ref{p:A.3.3} $H_1(\Sigma,\bar{\mathbb{S}}_{A'})$ 
is just the domain of the adjoint operator ${\cal D}^*$, we 
have that $\lambda^A\in\ker({\cal D})$ precisely when $\lambda
^A\in({\rm Im}({\cal D}^*))^\bot$, i.e. $\ker({\cal D})=({\rm 
Im}({\cal D}^*))^\bot$. Thus, by complex conjugation, we obtain 
from this that $\ker({\cal D}^*)=({\rm Im}({\cal D}))^\bot$, and 
hence that $({\rm Im}({\cal D}))^\bot$ is finite dimensional. 
Recalling that ${\rm Im}({\cal D})$ is closed, it is clear that 
$({\rm Im}({\cal D}))^\bot$ is isomorphic to $L_2(\Sigma,\bar{
\mathbb{S}}_{A'})/{\rm Im}({\cal D}):=\coker({\cal D})$. Since 
$\dim\ker({\cal D})=\dim\coker({\cal D})$, the index is 
vanishing.     \hfill $\Box$

\medskip

Proposition \ref{p:A.3.3} implies another important result, 
namely the following decomposition of the space $C^\infty(\Sigma,
\mathbb{S}^A)$ and its $L_2$-closure (yielding also the 
corresponding decomposition of their complex conjugate spaces): 

\begin{proposition}
\begin{equation}
L_2(\Sigma,\mathbb{S}^A)=\ker\bigl({\cal D}\bigr)\oplus{\rm Im}
\bigl({\cal D}^*\bigr), \hskip 20pt
C^\infty(\Sigma,\mathbb{S}^A)=\ker\bigl({\cal D}\bigr)\oplus
{\rm Im}\bigl({\cal D}^*\vert_{C^\infty}\bigr). \label{eq:A.3.2}
\end{equation}
\label{p:A.3.5}
\end{proposition}
{\it Proof:} Since $\dim({\cal D})$ is finite dimensional, it is 
closed even in $L_2(\Sigma,\mathbb{S}^A)$, and hence the orthogonal 
decomposition $L_2(\Sigma,\mathbb{S}^A)=\ker({\cal D})\oplus(\ker
({\cal D}))^\bot$ is well defined. Clearly, $\lambda^A\in\ker({\cal 
D})$ is equivalent to $\langle{\cal D}_{A'A}\lambda^A,\bar\omega_{A'}
\rangle=0$ for any $\bar\omega_{A'}\in L_2(\Sigma,\bar{\mathbb{S}}
_{A'})$. Since $H_1(\Sigma,\bar{\mathbb{S}}_{A'})\subset L_2(\Sigma,
\bar{\mathbb{S}}_{A'})$ is dense, $\lambda^A\in\ker({\cal D})$ is 
still equivalent to $0=\langle{\cal D}_{A'A}\lambda^A,\bar\mu_{A'}
\rangle=\langle\lambda^A,{\cal D}^{AA'}\bar\mu_{A'}\rangle$ for any 
$\bar\mu_{A'}\in H_1(\Sigma,\bar{\mathbb{S}}_{A'})$, i.e $\ker({\cal 
D})=({\rm Im}({\cal D}^*))^\bot$. Since ${\rm Im}({\cal D})\subset 
L_2(\Sigma,\bar{\mathbb{S}}_{A'})$ is closed and ${\cal D}^*$ is 
minus the complex conjugate of ${\cal D}$, this implies that 
$(\ker({\cal D}))^\bot=({\rm Im}({\cal D}^*))^{\bot\bot}=\overline{
{\rm Im}({\cal D}^*)}={\rm Im}({\cal D}^*)$. (Here overline denotes 
closure in the $L_2$-norm topology.) 

To prove the second decomposition, recall that the elements of 
$\ker({\cal D})$ are smooth, and let $\lambda^A\in C^\infty(\Sigma,
\mathbb{S}^A)$. Let $\lambda^A=\lambda^A_0+\lambda^A_1$ be the 
orthogonal decomposition corresponding to $L_2(\Sigma,\mathbb{S}
^A)=\ker({\cal D})\oplus{\rm Im}({\cal D}^*)$. Then $\lambda^A_1=
\lambda^A-\lambda^A_0$ is smooth, and there exists $\bar\mu_{A'}
\in H_1(\Sigma,\bar{\mathbb{S}}_{A'})$ such that $\lambda^A_1=
{\cal D}^{AA'}\bar\mu_{A'}$. Since ${\cal D}^{AA'}\bar\mu_{A'}$ is 
smooth, it belongs to $H_k(\Sigma,\bar{\mathbb{S}}_{A'})$ for any 
$k\in\mathbb{N}$. Thus, by $\bar\mu_{A'}\in H_1(\Sigma,\bar
{\mathbb{S}}_{A'})$ and Lemma \ref{l:A.2.2}, it follows that $\bar
\mu_{A'}\in H_k(\Sigma,\bar{\mathbb{S}}_{A'})$ for any $k\in
\mathbb{N}$, and hence, by the Sobolev lemma (Theorem 
\ref{th:A.1.1}.2), that $\bar\mu_{A'}\in C^\infty(\Sigma,\bar
{\mathbb{S}}_{A'})$. \hfill $\Box$

\medskip

Following the general rule (see \cite{Ka}, pp 167), the domain 
${\rm Dom}({\cal T}^*)$ of the adjoint of the 3-surface twistor 
operator is defined by 

\begin{eqnarray}
{\rm Dom}\bigl({\cal T}^*\bigr):=\bigl\{\,\phi_{ABC}\in L_2(\Sigma,
 \mathbb{S}_{(ABC)})\,\!\!\!\!&\vert\!\!\!\!&\,\exists\nu_A\in L_2
 (\Sigma, \mathbb{S}_A): \label{eq:A.3.3} \\
\langle{\cal D}_{(AB}\lambda_{C)},\phi_{ABC}\rangle\!\!\!\!&=\!\!\!\!
 &\langle\lambda_A,\nu_A\rangle \,\,\,\forall\,\lambda_A\in H_1(
 \Sigma,\mathbb{S}_A)\,\bigr\}. \nonumber
\end{eqnarray}
(Note that $\nu_A$ here is necessarily orthogonal to $\ker({\cal 
T})$.) Repeating the first part of the proof of Proposition 
\ref{p:A.3.3} we can see at once that $H_1(\Sigma,\mathbb{S}
_{(ABC)})\subset{\rm Dom}({\cal T}^*)$, and hence ${\rm Dom}({\cal 
T}^*)$ is dense in $L_2(\Sigma,\mathbb{S}_A)$. (Moreover, it is 
clear that this ${\cal T}^*$ is the extension of the formal adjoint 
of ${\cal T}$ introduced in subsection \ref{sub-4.3}.) However, the 
proof of the inclusion in the opposite direction fails, because we 
do not have a fundamental elliptic type estimate for ${\cal T}^*$ 
(that we did have for ${\cal T}$). 

Since the cokernel of ${\cal T}$ is infinite dimensional, it is not 
Fredholm. Similarly, ${\cal T}^*$ is not Fredholm either, because 
it has infinite dimensional kernel. However, we have that 

\begin{proposition}
${\cal T}^*:{\rm Dom}\bigl({\cal T}^*\bigr)\subset L_2(\Sigma,
\mathbb{S}_{(ABC)})\rightarrow L_2(\Sigma,\mathbb{S}_A)$ is a closed 
operator and ${\cal T}^{**}={\cal T}$. 
\label{p:A.3.6}
\end{proposition}
{\it Proof:} Since the defining equation of ${\rm Dom}({\cal T}^*)$ 
in (\ref{eq:A.3.3}) is just the condition that the inverse graph of 
$-{\cal T}^*$, defined by $G'(-{\cal T}^*):=\{(-{\cal T}^*\phi,\phi)
\,\vert\,\phi\in{\rm Dom}({\cal T}^*)\,\}\subset L_2(\Sigma,\mathbb{S}
_A)\times L_2(\Sigma,\mathbb{S}_{(ABC)})$, is the annihilator of the 
graph of ${\cal T}$, it is always closed. Hence, ${\cal T}^*$ is a 
closed operator (like every adjoint operator, see \cite{Ka}, pp. 168). 
${\cal T}^{**}={\cal T}$ follows from the reflexivity of the  $L_2$ 
spaces and the fact that ${\cal T}$ is closed.   \hfill $\Box$

\medskip
Although $\ker({\cal T}^*)$ is infinite dimensional, and we do not 
have an estimate for ${\cal T}^*$ analogous to (\ref{eq:A.2.6.b}), 
the kernel and the range of ${\cal T}^*$ are closed: 

\begin{proposition}
$\ker({\cal T}^*)\subset L_2(\Sigma,\mathbb{S}_{(ABC)})$ and ${\rm 
Im}({\cal T}^*)\subset L_2(\Sigma,\mathbb{S}_A)$ are closed 
subspaces. 
\label{p:A.3.7}
\end{proposition}
{\it Proof:} Let $\{\phi^i_{ABC}\}$, $i\in\mathbb{N}$, be a Cauchy 
sequence in $\ker({\cal T}^*)$ with respect to the $L_2$-norm, and 
let $\phi_{ABC}$ be its limit. Then ${\cal T}^*\phi^i_{ABC}=0$ for 
all $i\in\mathbb{N}$, i.e. both $\{\phi^i_{ABC}\}$ and $\{{\cal T}
^*\phi^i_{ABC}\}$ are Cauchy and ${\cal T}^*\phi^i_{ABC}\rightarrow0$. 
But since the operator ${\cal T}^*$ is closed, $\phi_{ABC}\in{\rm 
Dom}({\cal T}^*)$ and ${\cal T}^*\phi_{ABC}=0$ follow, implying that 
$\phi_{ABC}\in\ker({\cal T}^*)$. Thus $\ker({\cal T}^*)$ is closed. 
The statement that ${\rm Im}({\cal T}^*)$ is closed is a direct 
consequence of the general closed range theorem of Banach (see e.g. 
\cite{Yosida}, pp. 205), adapted to Hilbert spaces, and the fact 
that ${\cal T}$ is a densely defined closed operator.  \hfill $\Box$

\medskip

As a consequence of this proposition, we have decomposition 
theorems analogous to the first statement of Proposition 
\ref{p:A.3.5}: 

\begin{proposition}
\begin{equation}
L_2(\Sigma,\mathbb{S}_A)=\ker\bigl({\cal T}\bigr)\oplus{\rm Im}
\bigl({\cal T}^*\bigr), \hskip 20pt
L_2(\Sigma,\mathbb{S}_{(ABC)})={\rm Im}\bigl({\cal T}\bigr)\oplus
\ker\bigl({\cal T}^*\bigr). \label{eq:A.3.4}
\end{equation}
\label{p:A.3.8}
\end{proposition}
{\it Proof:} To prove the first, recall that $\ker({\cal T})$ 
is finite dimensional (Proposition \ref{p:A.3.1}), and hence 
it is closed in $L_2(\Sigma,\mathbb{S}_A)$. Hence we have the 
well defined $L_2$-orthogonal decomposition $L_2(\Sigma,
\mathbb{S}_A)=\ker({\cal T})\oplus(\ker({\cal T}))^\bot$, and 
we will show that $(\ker({\cal T}))^\bot=\overline{{\rm Im}
({\cal T}^*)}$, and hence by the previous proposition the 
statement follows. 
Thus, suppose that $\lambda_A\in\ker({\cal T})$. Then $0=
\langle{\cal D}_{(AB}\lambda_{C)},\phi_{ABC}\rangle=\langle
\lambda_A,{}^+{\cal D}^{BC}\phi_{ABC}\rangle$ for any $\phi
_{ABC}\in{\rm Dom}({\cal T}^*)$, i.e. $\ker({\cal T})\subset
({\rm Im}({\cal T}^*))^\bot$. Conversely, let $\lambda_A\in(
{\rm Im}({\cal T}^*))^\bot\cap H_1(\Sigma,\mathbb{S}_A)$. Then 
$0=\langle\lambda_A,{}^+{\cal D}^{BC}\phi_{ABC}\rangle=\langle
{\cal D}_{(AB}\lambda_{C)},\phi_{ABC}\rangle$ for any $\phi
_{ABC}\in{\rm Dom}({\cal T}^*)$. Since, however, ${\rm Dom}
({\cal T}^*)\subset L_2(\Sigma,\mathbb{S}_{(ABC)})$ is dense, 
this implies that ${\cal D}_{(AB}\lambda_{C)}=0$, i.e. 
$\lambda_A\in\ker({\cal T})$. Hence we obtained that $({\rm 
Im}({\cal T}^*))^\bot\cap H_1(\Sigma,\mathbb{S}_A)\subset
\ker({\cal T})\subset({\rm Im}({\cal T}^*))^\bot$. Taking 
its closure and recalling that $\ker({\cal T})\subset L_2
(\Sigma,\mathbb{S}_A)$ is closed and that $H_1(\Sigma,
\mathbb{S}_A)\subset L_2(\Sigma,\mathbb{S}_A)$ is dense, 
we obtain that $\ker({\cal T})=({\rm Im}({\cal T}^*))^\bot$, 
i.e. that $\overline{{\rm Im}({\cal T}^*)}=({\rm Im}({\cal 
T}^*))^{\bot\bot}=(\ker({\cal T}))^\bot$. 

The proof of the second decomposition is similar: Since 
${\rm Im}({\cal T})\subset L_2(\Sigma,\mathbb{S}_{(ABC)})$ 
is closed (Proposition \ref{p:A.3.2}), there is the 
decomposition $L_2(\Sigma,\mathbb{S}_{(ABC)})={\rm Im}(
{\cal T})\oplus({\rm Im}({\cal T}))^\bot$, and we prove that 
$({\rm Im}({\cal T}))^\bot=\overline{\ker({\cal T}^*)}$. If 
$\phi_{ABC}\in\ker({\cal T}^*)$, then $0=\langle\lambda_A,
{}^+{\cal D}^{BC}\phi_{ABC}\rangle=\langle{\cal D}_{(AB}\lambda
_{C)},\phi_{ABC}\rangle$ for any $\lambda_A\in H_1(\Sigma,
\mathbb{S}_A)$, and hence $\ker({\cal T}^*)\subset({\rm Im}
({\cal T}))^\bot$. Conversely, let $\phi_{ABC}\in({\rm Im}
({\cal T}))^\bot\cap{\rm Dom}({\cal T}^*)$. Then $0=\langle
{\cal D}_{(AB}\lambda_{C)},\phi_{ABC}\rangle=\langle\lambda_A,
{}^+{\cal D}^{BC}\phi_{ABC}\rangle$ for any $\lambda_A\in H_1
(\Sigma,\mathbb{S}_A)$. Since $H_1(\Sigma,\mathbb{S}_A)\subset 
L_2(\Sigma,\mathbb{S}_A)$ is dense, this implies that ${}^+
{\cal D}^{BC}\phi_{ABC}=0$, i.e. that $({\rm Im}({\cal T}))
^\bot\cap{\rm Dom}({\cal T}^*)\subset\ker({\cal T}^*)\subset
({\rm Im}({\cal T}))^\bot$. Recalling that ${\rm Dom}({\cal T}
^*)\subset L_2(\Sigma,\mathbb{S}_{(ABC)})$ is dense and that 
$\ker({\cal T}^*)$ is closed by the previous proposition, the 
closure of this line of inclusions yields that $({\rm Im}
({\cal T}))^\bot=\overline{\ker({\cal T}^*)}=\ker({\cal T}^*)$. 
\hfill $\Box$

\medskip
Since $\ker({\cal T})$ is finite dimensional, repeating the 
proof of the second statement in Proposition \ref{p:A.3.5} 
one can show that $C^\infty(\Sigma,\mathbb{S}_A)=\ker({\cal 
T})\oplus{\rm Im}({\cal T}^*\vert_{C^\infty})$ also holds.


\subsection{The second order operators ${\cal D}^*{\cal 
D}$ and ${\cal T}^*{\cal T}$}
\label{sub-A.4}

The domain of ${\cal D}^*{\cal D}$ and ${\cal T}^*{\cal T}$ will 
be defined, respectively, by 

\begin{align*}
&{\rm Dom}\bigl({\cal D}^*{\cal D}\bigr):=\bigl\{\,\lambda^A\in 
 H_1\bigl(\Sigma,\mathbb{S}^A\bigr)\,\vert\,{\cal D}_{A'A}\lambda
 ^A\in{\rm Dom}({\cal D}^*)\,\bigr\}, \\
&{\rm Dom}\bigl({\cal T}^*{\cal T}\bigr):=\bigl\{\lambda_A\in 
H_1\bigl(\Sigma,\mathbb{S}_A\bigr)\,\vert\,{\cal D}_{(AB}
\lambda_{C)}\in{\rm Dom}\bigl({\cal T}^*\bigr)\,\bigr\},
\end{align*}
where ${\rm Dom}({\cal D}^*)=H_1(\Sigma,\bar{\mathbb{S}}_{A'})$ 
(see Proposition \ref{p:A.3.3}) and $H_1(\Sigma,\mathbb{S}_{(ABC)})
\subset{\rm Dom}({\cal T}^*)$.

\begin{lemma}
\begin{align}
&\ker({\cal D}^*{\cal D})=\ker({\cal D}), \hskip 20pt
\ker({\cal T}^*{\cal T})=\ker({\cal T}); \label{eq:A.4.1a} \\
&{\rm Im}({\cal D}^*)={\rm Im}({\cal D}^*{\cal D}), \hskip 20pt
{\rm Im}({\cal T}^*)={\rm Im}({\cal T}^*{\cal T}). 
\label{eq:A.4.1b}
\end{align}
\label{l:A.4.0}
\end{lemma}
{\it Proof:} The inclusion $\ker({\cal D})\subset\ker({\cal D}
^*{\cal D})$ is obviously true. To prove the inclusion in the 
opposite direction, suppose that $\lambda^A\in\ker({\cal D}^*
{\cal D})$. Then $0=\langle{\cal D}^{AA'}{\cal D}_{A'B}\lambda^B,
\lambda^A\rangle=\langle{\cal D}_{A'A}\lambda^A,{\cal D}_{A'B}
\lambda^B\rangle$, i.e. ${\cal D}_{A'A}\lambda^A=0$. Hence $\ker
({\cal D}^*{\cal D})\subset\ker({\cal D})$. The proof for the 
3-surface twistor operator is similar. 

To prove (\ref{eq:A.4.1b}) for the Sen--Witten operator we should 
use the first of the decompositions (\ref{eq:A.3.2}), and for the 
3-surface twistor operator we should use the first of 
(\ref{eq:A.3.4}). Since the two proofs are similar, we give the 
proof only for the first: 

\begin{eqnarray*}
{\rm Im}\bigl({\cal D}^*\bigr)\!\!\!\!&=\!\!\!\!&\{{\cal D}^{AA'}
 \bar\mu_{A'}\vert\,\bar\mu_{A'}\in H_1(\Sigma,\bar{\mathbb{S}}_{A'}
 )\}= \\
\!\!\!\!&=\!\!\!\!&\{{\cal D}^{AA'}\bar\mu_{A'}\vert\,\bar\mu_{A'}
 \in H_1(\Sigma,\bar{\mathbb{S}}_{A'})\cap\big(\ker({\cal D}^*)
 \oplus{\rm Im}({\cal D}))\}=\\
\!\!\!\!&=\!\!\!\!&\{{\cal D}^{AA'}\bar\mu_{A'}\vert\,\bar\mu_{A'}
 \in H_1(\Sigma,\bar{\mathbb{S}}_{A'})\cap{\rm Im}({\cal D})\}=\\
\!\!\!\!&=\!\!\!\!&\{{\cal D}^{AA'}{\cal D}_{A'B}\omega^B\vert\,
 \omega^A\in H_1(\Sigma,\mathbb{S}^A),\, {\cal D}_{A'A}\omega^A\in 
 H_1(\Sigma,\bar{\mathbb{S}}_{A'})\}={\rm Im}\bigl({\cal D}^*{\cal 
 D}\bigr).
\end{eqnarray*}

\hfill $\Box$

An immediate consequence of this lemma is the decomposition $L_2(
\Sigma,\mathbb{S}_A)=\ker({\cal D}^*{\cal D})\oplus{\rm Im}({\cal 
D}^*{\cal D})=\ker({\cal T}^*{\cal T})\oplus{\rm Im}({\cal T}^*
{\cal T})$, where all the subspaces are closed.

One of the key properties of the second order operators is their 
Fredholm property:

\begin{proposition}
${\cal D}^*{\cal D}$ and ${\cal T}^*{\cal T}$ are positive, 
self-adjoint Fredholm operators. 
\label{p:A.4.1}
\end{proposition}
{\it Proof:} 
{\em Positivity:} Since both ${\cal D}^*{\cal D}$ and ${\cal T}
^*{\cal T}$ are positive operators on the space of the smooth 
spinor fields, which is dense in their domain, the operators are 
positive on ${\rm Dom}({\cal D}^*{\cal D})$ and ${\rm Dom}({\cal 
T}^*{\cal T})$, respectively, too. 

{\em Self-adjointness}: We should show that ${\rm Dom}({\cal D}^*
{\cal D})={\rm Dom}(({\cal D}^*{\cal D})^*)$ and ${\rm Dom}({\cal 
T}^*{\cal T})={\rm Dom}(({\cal T}^*{\cal T})^*)$. Since the two 
proofs are similar, we prove it only for ${\cal T}^*{\cal T}$. 

Suppose that $\chi_A\in{\rm Dom}({\cal T}^*{\cal T})$, i.e. 
$\chi_A\in H_1(\Sigma,\mathbb{S}_A)$ such that ${\cal D}_{(AB}
\chi_{C)}\in{\rm Dom}({\cal T}^*)$. Thus there exists a spinor 
field $\nu_A\in L_2(\Sigma,\mathbb{S}_A)$ such that $\langle
{\cal D}_{(AB}\lambda_{C)},{\cal D}_{(AB}\chi_{C)}\rangle=\langle
\lambda_A,\nu_A\rangle$ for any $\lambda_A\in{\rm Dom}({\cal T})=
H_1(\Sigma,\mathbb{S}_A)$. 
Since, however, ${\rm Dom}({\cal T}^*{\cal T})\subset{\rm Dom}(
{\cal T})$, we have that $\langle{}^+{\cal D}^{BC}{\cal D}_{(AB}
\lambda_{C)},\chi_A\rangle=\langle{\cal D}_{(AB}\lambda_{C)},{\cal 
D}_{(AB}\chi_{C)}\rangle=\langle\lambda_A,\nu_A\rangle$ for any 
$\lambda_A\in{\rm Dom}({\cal T}^*{\cal T})$. Thus $\chi_A\in{\rm 
Dom}(({\cal T}^*{\cal T})^*)$, i.e. ${\rm Dom}({\cal T}^*{\cal 
T})\subset{\rm Dom}(({\cal T}^*{\cal T})^*)$. 

Conversely, suppose that $\chi_A\in{\rm Dom}(({\cal T}^*{\cal T})
^*)$. Then there exists a spinor field $\nu_A\in(\ker({\cal T}^*
{\cal T}))^\bot$ such that $\langle{}^+{\cal D}^{BC}{\cal D}_{(AB}
\lambda_{C)},\chi_A\rangle=\langle\lambda_A,\nu_A\rangle$ for any 
$\lambda_A\in{\rm Dom}({\cal T}^*{\cal T})$. However, by 
(\ref{eq:A.4.1a}) and the first decomposition in (\ref{eq:A.3.4}) 
$(\ker({\cal T}^*{\cal T}))^\bot=(\ker({\cal T}))^\bot={\rm Im}(
{\cal T}^*)$, and hence $\nu_A={}^+{\cal D}^{BC}\phi_{ABC}$ for some 
$\phi_{ABC}\in{\rm Dom}({\cal T}^*)$. On the other hand, by the 
second decomposition in (\ref{eq:A.3.4}), we have that ${\rm Dom}
({\cal T}^*)={\rm Dom}({\cal T}^*)\cap(\ker({\cal T}^*)\oplus{\rm 
Im}({\cal T}))$, and therefore $\phi_{ABC}=\phi^0_{ABC}+{\cal D}
_{(AB}\omega_{C)}$ for some $\omega_A\in{\rm Dom}({\cal T})=H_1(
\Sigma,\mathbb{S}_A)$ and $\phi^0_{ABC}\in\ker({\cal T}^*)$. Hence, 
$\nu_A={}^+{\cal D}^{BC}{\cal D}_{(AB}\omega_{C)}$, by means of which 
$\langle{}^+{\cal D}^{BC}{\cal D}_{(AB}\lambda_{C)},\chi_A\rangle=
\langle\lambda_A,\nu_A\rangle=\langle{}^+{\cal D}^{BC}{\cal D}_{(AB}
\lambda_{C)},\omega_A\rangle$, i.e. $\langle{}^+{\cal D}^{BC}{\cal 
D}_{(AB}\lambda_{C)},\chi_A-\omega_A\rangle=0$. Thus $\omega^0_A:=
\chi_A-\omega_A\in({\rm Im}({\cal T}^*{\cal T}))^\bot=({\rm Im}
({\cal T}^*))^\bot=\ker({\cal T})\subset C^\infty(\Sigma,\mathbb{S}
_A)$, where we used (\ref{eq:A.4.1b}), (\ref{eq:A.3.4}) and 
Proposition \ref{p:A.3.1}. Consequently, $\chi_A=\omega_A+\omega
^0_A\in H_1(\Sigma,\mathbb{S}_A)$ and $\chi_A\in{\rm Dom}({\cal 
T}^*)$, i.e. $\chi_A\in{\rm Dom}({\cal T}^*{\cal T})$. Therefore, 
${\rm Dom}(({\cal T}^*{\cal T})^*)\subset{\rm Dom}({\cal T}^*
{\cal T})$. 

{\em The Fredholm property}: We prove this only for ${\cal T}^*
{\cal T}$, the proof for ${\cal D}^*{\cal D}$ is similar. By 
Proposition \ref{p:A.3.1} and (\ref{eq:A.4.1a}) $\ker({\cal T}^*
{\cal T})=\ker({\cal T})$ is finite dimensional. By 
(\ref{eq:A.4.1b}) we have that $\coker({\cal T}^*{\cal T}):=L_2(
\Sigma,\mathbb{S}_A)/{\rm Im}({\cal T}^*{\cal T})=(\ker({\cal T})
\oplus{\rm Im}({\cal T}^*))/{\rm Im}({\cal T}^*)\approx\ker({\cal 
T})$, which is also finite dimensional. By Proposition 
\ref{p:A.3.7} ${\rm Im}({\cal T}^*)$ is closed, and hence by 
(\ref{eq:A.4.1b}) ${\rm Im}({\cal T}^*{\cal T})$ is also closed. 
Therefore, ${\cal T}^*{\cal T}$ is Fredholm.   \hfill $\Box$
\medskip

Finally, we clarify the spectral properties of $2{\cal D}^*{\cal D}$ 
and $2{\cal T}^*{\cal T}$. Thus let $E_{\alpha^2}$ and $E_{\tau^2}$ 
denote the space of their eigenspinors with eigenvalue $\alpha^2$ 
and $\tau^2$, respectively. Adapting an analogous theorem of 
\cite{FaSc} to the present operators, we have our last statement: 

\begin{proposition}
The resolvent operators of $2{\cal D}^*{\cal D}$ and $2{\cal T}^*
{\cal T}$ are compact. 
\label{p:A.4.2}
\end{proposition}
{\it Proof:} We prove the statement only for $2{\cal T}^*{\cal T}$. 
The proof for $2{\cal D}^*{\cal D}$ is similar. 
First we show that $2{\cal T}^*{\cal T}$ does not have any 
eigenvalue in the interval $(0,2/{C'}^2_0)$, where $C'_0$ is the 
positive constant in Lemma \ref{l:A.2.3}. 

For, let $\tau^2>0$ and suppose that $\lambda_A\in E_{\tau^2}$. 
Then by $\lambda_A\in{\rm Dom}({\cal T}^*{\cal T})$ and $\ker
({\cal T}^*{\cal T})=\ker({\cal T})$ we have that $\lambda_A\in
(\ker({\cal T}))^\bot\cap H_1(\Sigma,\mathbb{S}_A)$, and hence by 
Lemma \ref{l:A.2.3} $\frac{1}{2}\tau^2\Vert\lambda_A\Vert^2_{L_2}=
\langle\frac{1}{2}\tau^2\lambda_A,\lambda_A\rangle=\langle{}^+
{\cal D}^{BC}{\cal D}_{(AB}\lambda_{C)},\lambda_A\rangle=\langle
{\cal D}_{(AB}\lambda_{C)},{\cal D}_{(AB}\lambda_{C)}\rangle=\Vert
{\cal D}_{(AB}\lambda_{C)}\Vert^2_{L_2}$ $\geq(1/C'_0)^2\Vert\lambda
_A\Vert^2_{H_1}\geq(1/C'_0)^2\Vert\lambda_A\Vert^2_{L_2}$, i.e. 
$\tau^2{C'}^2_0\Vert\lambda_A\Vert^2_{L_2}\geq2\Vert\lambda_A\Vert
^2_{L_2}$. If, however, $\tau^2{C'}^2_0<2$, then this implies that 
$\lambda_A=0$, i.e. that $E_{\tau^2}=\emptyset$. 

Next, let us define the operator $\Delta:=2{\cal T}^*{\cal T}-
\tau^2I:{\rm Dom}({\cal T}^*{\cal T})\rightarrow L_2(\Sigma,
\mathbb{S}_A)$, which is continuous and, by $E_{\tau^2}=\emptyset$, 
$\ker(\Delta)=\emptyset$. (Here $I$ denotes the identity operator.) 
Let us define $W:=({\rm Im}(\Delta))^\bot\subset L_2(\Sigma,
\mathbb{S}_A)$, and suppose that this is not empty. Then, if 
$\omega_A\in W$, it follows that 

\begin{equation*}
0=\langle\Delta_A{}^B\lambda_B,\omega_A\rangle=\langle2{}^+{\cal 
D}^{BC}{\cal D}_{(AB}\lambda_{C)}-\tau^2\lambda_A,\omega_A\rangle
\end{equation*}
for any $\lambda_A\in{\rm Dom}(\Delta)={\rm Dom}({\cal T}^*{\cal 
T})$, i.e. $\langle{}^+{\cal D}^{BC}{\cal D}_{(AB}\lambda_{C)},
\omega_A\rangle=\langle\lambda_A,\frac{1}{2}\tau^2\omega_A\rangle$ 
holds. However, this means that $W\subset{\rm Dom}(({\cal T}^*
{\cal T})^*)={\rm Dom}({\cal T}^*{\cal T})$ and ${}^+{\cal D}^{BC}
{\cal D}_{(AB}\omega_{C)}=\frac{1}{2}\tau^2\omega_A$, which 
contradicts $E_{\tau^2}=\emptyset$. Hence $W=\emptyset$, i.e. ${\rm 
Im}(\Delta)=L_2(\Sigma,\mathbb{S}_A)$. 

Thus, $\Delta:{\rm Dom}({\cal T}^*{\cal T})\rightarrow L_2
(\Sigma,\mathbb{S}_A)$ is a continuous bijection. But then by 
the open mapping theorem (see e.g. \cite{Pa}, pp 107) $\Delta$ 
is a topological vector space isomorphism, admitting a 
continuous inverse $\Delta^{-1}:L_2(\Sigma,\mathbb{S}_A)
\rightarrow{\rm Dom}({\cal T}^*{\cal T})\subset H_1(\Sigma,
\mathbb{S}_A)$. Therefore, there exists a positive constant 
$K_{\tau^2}$ such that $\Vert\Delta^{-1}_A{}^B\omega_B\Vert_{H_1}
\leq K_{\tau^2}\Vert\omega_A\Vert_{L_2}$ for any $\omega_A\in L_2
(\Sigma,\mathbb{S}_A)$. Hence, if $\{\omega^i_A\}$, $i\in
\mathbb{N}$, is a bounded sequence in $L_2(\Sigma,\mathbb{S}
_A)$, then $\{\Delta^{-1}_A{}^B\omega^i_A\}$ is bounded in $H_1
(\Sigma,\mathbb{S}_A)$. But by the Rellich lemma the inclusion 
$H_1(\Sigma,\mathbb{S}_A)\subset L_2(\Sigma,\mathbb{S}_A)$ is 
compact, implying that there is a subsequence $\{\Delta^{-1}_A
{}^B\omega^{i_k}_A\}$, $k\in\mathbb{N}$, which is convergent in 
$L_2(\Sigma,\mathbb{S}_A)$. Therefore, the resolvent $\Delta
^{-1}$ of $2{\cal T}^*{\cal T}-\tau^2I$, as a bounded linear 
operator $L_2(\Sigma,\mathbb{S}_A)\rightarrow L_2(\Sigma,
\mathbb{S}_A)$, is compact. \hfill $\Box$
\medskip

Applying the results on the spectral properties of compact 
operators (see e.g. \cite{Pa}, Theorem 16, pp 114) to the 
resolvent, and recalling how the spectra of the operator and its 
resolvent are related to each other (see e.g. \cite{Ka}, pp 187), 
we obtain (see e.g. \cite{LaMi}, pp 196) that 
(1) the spectrum of $2{\cal D}^*{\cal D}$ is purely discrete 
with the only accumulation point at infinity, 
(2) there is a positive constant $c$ such that for the $k$th 
eigenvalue $\alpha^2_k\geq ck^{(1/6)}$, 
(3) the space $E_{\alpha^2}$ of the eigenspinors with eigenvalue 
$\alpha^2$ is finite dimensional and the eigenspinors are smooth, 
(4) the spaces $E_{\alpha^2}$, $E_{\beta^2}$ with $\alpha^2\not=\beta
^2$ are orthogonal to each other, 
(5) $L_2(\Sigma,\mathbb{S}^A)=\oplus_{\alpha^2\in[0,\infty)}E
_{\alpha^2}$. 
(For the proof see e.g. \cite{Pa,Ka,LaMi}. However, we note that 
the bound for the growth of rate of the eigenvalues given in 
\cite{LaMi}, pp. 196, can be increased, and the bound given in 
(2) above is this greater one.) Clearly $2{\cal T}^*{\cal T}$ has 
similar spectral properties. 


\hskip 25pt

The author is grateful to Florian Beyer, J\"org Frauendiener and P\'eter 
Vecserny\'es for the discussions on various parts of the present paper; 
and special thanks to Juan A. Valiente-Kroon for the discussions on his 
related results and ideas. This work was partially supported by the 
Hungarian Scientific Research Fund (OTKA) grant K67790.


\noindent

\begin{thebibliography}{9999}


\bibitem{Wi} E. Witten, A new proof of the positive energy theorem, 
          Commun. Math. Phys. {\bf 30} 381--402 (1981)

\bibitem{Ne} J. M. Nester, A new gravitational energy expression and 
          with a simple positivity proof, Phys. Lett. A, {\bf 83} 
          241--242 (1981)

\bibitem{GHHP} G. W. Gibbons, S. W. Hawking, G. T. Horowitz, M. J. 
          Perry, Positive mass theorem for black holes, Commun. Math. 
          Phys. {\bf 88} 295--308 (1983)

\bibitem{ReTo} O. Reula, K. P. Tod, Positivity of the Bondi energy, 
         J. Math. Phys. {\bf 25} 1004--1008 (1984) 

\bibitem{Pe} R. Penrose, Naked singularities, Ann. N. Y. Acad. Sci. 
         {\bf 224} 125-134 (1973)

\bibitem{Mars} M. Mars, Present status of the Penrose inequality , 
         Class. Quantum Grav. {\bf 26} 193001 (2009), arXiv: 
         0906.5566  

\bibitem{BaKr} T. B\"ackdahl, J. A. V. Kroon, Approximate twistors 
         and positive mass, Class. Quantum Grav. {\bf 28} 075010 (2011), 
         arXiv:1011.3712 [gr-qc]

\bibitem{Tod} K. P. Tod, Three-surface twistors and conformal 
         embedding, Gen. Rel. Grav. {\bf 16} 435--443 (1984) 

\bibitem{Sz09} L. B. Szabados, Quasi-local energy-momentum and 
         angular momentum in GR: A review article, Living Rev. 
         Relativity {\bf 12} (2009) 4, {\tt 
         http://www.livingreviews.org/ lrr-2009-4} 

\bibitem{Parker} T. H. Parker, Gauge choice in Witten's energy 
          expression, Commun. Math. Phys. {\bf 100} 471--480 (1985) 

\bibitem{Ne1}
         J. M. Nester, A positive gravitational energy proof, Phys.
         Lett. A {\bf 139} 112-114 (1989)
\bibitem{Ne2}
         J. M. Nester, A gauge condition for orthonormal three-frames,
          J. Math. Phys. {\bf 30} 624--626 (1989)
\bibitem{Ne3}
         J. M. Nester, Special orthonormal frames, J. Math. Phys.
          {\bf 33} 910--913 (1992)


\bibitem{Yau} S.-T. Yau, Problem section, in {\it Seminar on 
         Differential Geometry}, pp. 669, Ann. Math. Studies No 102, 
         Princeton University Press, Princeton 1982

\bibitem{L} A. Lichnerowicz, Spineurs harmoniques, C. R. Acad. Sci. 
          Paris A--B {\bf 257} 7--9 (1963)

\bibitem{TFr} T. Friedrich, Der erste Eigenwert des Dirac-operators 
         einer kompakten Riemannschen Mannigfaltigkeit nichtnegativer 
         Skalarkr\"ummung, Math. Nachr. {\bf 97} 117--146 (1980)

\bibitem{Hi86} O. Hijazi, A conformal lower bound for the smallest 
         eigenvalue of the Dirac operator and Killing spinors, Commun. 
         Math. Phys. {\bf 104} 151--162 (1986)

\bibitem{Hi95} O. Hijazi, Lower bounds for the eigenvalues of the 
         Dirac operator, J. Geom. Phys. {\bf 16} 27--38 (1995) 

\bibitem{TF00} T. Friedrich, {\it Dirac Operators in Riemannian 
         Geometry}, Graduate Studies in Mathematics, Vol 25, AMS 
         Providence, Rhode Island 2000

\bibitem{FaSc} S. Farinelli, G. Schwarz, On the spectrum of the 
         Dirac operators under boundary conditions, J. Geom. Phys. 
         {\bf 28} 67--87 (1998)

\bibitem{Ba92} C. B\"ar, Lower eigenvalue estimates for Dirac 
         operators, Math. Ann. {\bf 293} 39--46 (1992) 

\bibitem{HiZh} O. Hijazi, X. Zhang, The Dirac--Witten operators on 
          spacelike hypersurfaces, Commun. Anal. Geom. {\bf 11} 
          737--750 (2003)


\bibitem{Se} A. Sen, On the existence of neutrino `zero-modes' in 
         vacuum spacetimes, J. Math. Phys. {\bf 22} 1781--1786 (1981) 


\bibitem{PRI}
         R. Penrose, W. Rindler, {\it Spinors and Spacetime}, Vol 1, 
          Cambridge University Press, Cambridge 1984

\bibitem{Re} O. Reula, Existence theorem for solutions of Witten's 
          equation and nonnegativity of total mass, J. Math. Phys. 
          {\bf 23} 810--814 (1982)
\bibitem{Fr} J. Frauendiener, Triads and the Witten equation, Class. 
           Quantum Gravity, {\bf 8} 1881--1187 (1991) 


\bibitem{Sz93} L. B. Szabados: On the positivity of the quasi-local 
         mass, Class. Quantum Grav. {\bf 10} 1899-1905 (1993)
\bibitem{Sz94} L. B. Szabados: Two dimensional Sen connections and 
         quasi-local energy-mom\-entum, Class. Quantum Grav. {\bf 
         11} 1847-1866 (1994), gr-qc/9402005

\bibitem{Sz96} L. B. Szabados: Quasi-local energy-momentum and 
         two-surface characterization of the {\it pp}-wave 
         spacetimes, Class. Quantum Grav. {\bf 13} 1661-1678 
         (1996), gr-qc/9512013

\bibitem{Sz94a} L. B. Szabados: Two dimensional Sen connections in 
         general relativity, Class. Quantum Grav. {\bf 11} 1833--1846 
         (1994), gr-qc/9402001

\bibitem{St} N. Steenrod, {\it The Topology of Fiber Bundles}, 
         Princeton University Press, Princeton 1951


\bibitem{PRII}
         R. Penrose, W. Rindler, {\it Spinors and Spacetime}, Vol 2, 
          Cambridge University Press, Cambridge 1986


\bibitem{FNSz} J. Frauendiener, J. M. Nester, L. B. Szabados, Witten 
         spinors on maximal, conformally flat hypersurfaces, Class. 
         Quantum Grav. {\bf 28} (2011) 185004, arXiv: 1105.5008 
         [gr-qc]

\bibitem{Wald} R. M. Wald, {\it General Relativity}, The University of 
         Chicago Press, Chicago 1984


\bibitem{Besse} A. L. Besse, {\it Einstein Manifolds}, Springer, 
            Berlin 2008

\bibitem{Adams:Fournier} R. A. Adams, J. J. F. Fournier, {\it Sobolev 
           Spaces}, Academic Press, Amsterdam 2003
\bibitem{Evans} L. C. Evans, {\it Partial Differential Equations}, 
           Graduate Studies in Mathematics No 19, AMS, Providence 
           RI 2002

\bibitem{Pa} R. Palais, {\it Seminar on the Atiyah--Singer Index 
         Theorem}, Princeton Univ. Press, Princeton 1965

\bibitem{Ka} T. Kato, {\it Perturbation Theory for Linear Operators}, 
           Classics in Mathematics Series, Springer, Berlin 1995
\bibitem{Yosida} K. Yosida, {\it Functional Analysis}, Springer, Berlin 
            1995

\bibitem{LaMi} H. B. Lawson, M.-L. Michelsohn, {\it Spin Geometry}, 
         Princeton University Press, Princeton 1989

\end{thebibliography}
\end{document}